\documentclass[a4paper,11pt]{article}

\usepackage{jheppub} 
\usepackage[utf8]{inputenc}
\usepackage{hyperref}

\definecolor{redcustom}{rgb}{0.8, 0.024, 0.302}
\definecolor{editGreen}{rgb}{0.267, 0.49, 0.18}

\definecolor{ValenciaBlue}{rgb}{0,0.171875,0.3203125}
\definecolor{ValenciaBlue1}{rgb}{0.408, 0.631, 0.835}
\definecolor{ValenciaBlue2}{rgb}{0.137, 0.306, 0.514}
\definecolor{ValenciaBlue3}{rgb}{0.259, 0.427, 0.655}
\definecolor{ValenciaPhysics}{rgb}{0.129, 0.114, 0.333}

\hypersetup{
     colorlinks=true,
     linkcolor=redcustom,
     filecolor=blue,
     citecolor = redcustom,      
     urlcolor=redcustom,
     }
\usepackage{booktabs}
\usepackage[T1]{fontenc}
\usepackage{siunitx}
\usepackage{xcolor,graphicx}
\usepackage{dcolumn}
\usepackage{bm}
\usepackage[normalem]{ulem}
\usepackage{braket}
\usepackage{bbold}
\usepackage{amsmath}
\usepackage{cancel}
\usepackage{slashed}
\usepackage{multicol}
\usepackage{multirow}
\usepackage{cleveref}
\usepackage{arydshln}
\usepackage{lscape}
\usepackage{pdflscape}
\usepackage{makecell}
\usepackage{graphicx}
\usepackage{subfigure}
\usepackage[export]{adjustbox}
\usepackage{tikz}
\usepackage[compat=1.1.0]{tikz-feynman}
\usepackage[compat=1.1.0]{tikz-feynhand}
\usepackage{dsfont}
\usepackage{placeins}

\newcommand{\sig}{\sigma}
\newcommand{\lzm}{\left(}
\newcommand{\dzm}{\right)}
\newcommand{\lzs}{\left[}
\newcommand{\dzs}{\right]}

\newcommand{\cL}{\mathcal{L}}
\newcommand{\cF}{F}
\newcommand{\cO}{\mathcal{O}}

\newcommand{\cM}{{\mathcal M}}
\newcommand{\cN}{{\mathcal N}}

\newcommand{\cC}{{\mathcal C}}

\newcommand{\mev}{\mathrm{MeV}}
\newcommand{\gev}{\mathrm{GeV}}

\newcommand{\re}{{\mathrm{Re}} \,}
\newcommand{\im}{{\mathrm{Im}} \,}
\newcommand{\hermc}{\text{h.c.}}

\newcommand{\Tr}{\mathop{\mathrm{Tr}}}
\newcommand{\tr}{\mathop{\mathrm{Tr}}}
\newcommand{\eminus}{\vcenter{\hbox{\scalebox{0.6}[1]{$ - $}}}}	%Narrow minus sign (for e.g. negative exponents)

\newcommand{\sscript}[1]{{\scriptscriptstyle \mathrm{#1}}}

\definecolor{deepskyblue}{rgb}{0.0, 0.75, 1.0}
\definecolor{aqua}{rgb}{0.0, 1.0, 1.0}
\definecolor{bronze}{rgb}{0.8, 0.5, 0.2}
\definecolor{electricyellow}{rgb}{1.0, 1.0, 0.0}
\definecolor{goldenyellow}{rgb}{1.0, 0.87, 0.0}
\definecolor{glaucous}{rgb}{0.38, 0.51, 0.71}
\colorlet{blueRef}{blue!80!black}
\colorlet{CodeColor}{red!70!black}

\title{
EFT for Neutrino Oscillations:\\Theory Developments and Application to JUNO
}

\author[]{Mart\'in Gonz\'alez-Alonso,}
\author[]{Ajdin Palavri\'c,}
\author[]{Suraj Prakash}

\affiliation[]{Instituto de F\'isica Corpuscular (IFIC), CSIC-Universitat de Val\`encia (UV), Spain}

\emailAdd{martin.gonzalez@ific.uv.es}
\emailAdd{ajdin.palavric@ific.uv.es}
\emailAdd{suraj.prakash@ific.uv.es}

\begin{document}

%%%%%%%%%%%%%%%%%%%%%%%%%%%%%%%%%%%%%%%%%%%%%%%%%%%%%%%%%%%%%%%%%%%%%%%%%%%%%%%%
\abstract
{We contribute to the systematic analysis of New Physics effects in neutrino experiments using Effective Field Theory (EFT) methods. 
We review and extend the quantum field-theoretical formalism for generic neutrino interactions, discussing the inclusion of matter effects and deriving the connection with the density matrix formalism. 
On the phenomenological side, we apply this framework to medium-baseline reactor neutrino experiments. We derive analytical expressions for the relevant oscillation observables and, for the first time, perform an EFT analysis of the recent JUNO dataset, extracting bounds on the leading non-standard interaction parameters.
}
%%%%%%%%%%%%%%%%%%%%%%%%%%%%%%%%%%%%%%%%%%%%%%%%%%%%%%%%%%%%%%%%%%%%%%%%%%%%%%%%

\maketitle

%%%%%%%%%%%%%%%%%%%%%%%%%%%%%%%%%%%%%%%%%%%%%%%%%%%%%%%%%%%%%%%%%%%%%%%%%%%%%%%%
\clearpage
\section{Introduction}
%%%%%%%%%%%%%%%%%%%%%%%%%%%%%%%%%%%%%%%%%%%%%%%%%%%%%%%%%%%%%%%%%%%%%%%%%%%%%%%%

Neutrino physics has been steadily evolving into a precision field. Experiments have long moved past establishing neutrino oscillations as a phenomenon, and are now sensitive to subleading effects that probe the boundaries of the Standard Model (SM). This transition is nicely illustrated by the recent results from the JUNO collaboration~\cite{JUNO:2025gmd}, which demonstrate the remarkable level of experimental control that has become achievable in this sector. As measurements become more precise, the (dis)agreement among the various experimental results itself becomes a non-trivial source of information: the degree to which data from different experiments can be reconciled within a common framework places increasingly stringent restrictions on New Physics (NP) extensions of the SM.
A particularly well-suited language for this kind of analysis is that of Effective Field Theory (EFT). When searching for NP with low-energy precision experiments, the EFT approach offers a systematic and model-independent way to parametrize deviations from SM predictions. Depending on the assumptions made about the relevant energy scales and degrees of freedom, one can follow an "EFT ladder", connecting a hierarchy of effective theories that bridge the gap between low-energy observables and the underlying high-energy physics. Importantly, the model-independent nature of the EFT approach demands the inclusion of all operators allowed at a given order. In the context of neutrino experiments, this implies the simultaneous inclusion of NP contributions to both neutral-current (NC) and charged-current (CC) interactions, a point that is especially important in the Standard Model EFT (SMEFT) context, where NC and CC operators are deeply intertwined by the $\mathrm{SU}(2)_L$ structure of the lepton doublet.
This EFT approach to neutrino oscillations was pursued in Ref.~\cite{Falkowski:2019xoe,Falkowski:2019kfn}, where a well-defined quantum field-theoretical (QFT) formalism was developed to treat, for the first time, generic CC interactions in oscillation phenomena. The formalism has since been extended and applied in a number of phenomenological studies~\cite{Falkowski:2021bkq,Du:2021rdg,Breso-Pla:2023tnz,Cherchiglia:2023aqp,Kopp:2024yvh,Coloma:2024ict,Breso-Pla:2025pds,Kling:2025zsb,Kopp:2025ffx}.

The present work contributes to this ongoing program on two fronts. On the theoretical side, we introduce a compact matrix notation that is physically illuminating, which allows a natural connection with the density matrix formalism; we further discuss the inclusion of matter effects in the presence of generic EFT interactions in the QFT framework. On the phenomenological side, taking the recent JUNO measurements as our main motivation, we study the medium-baseline reactor neutrino experiments within this framework. We derive analytical expressions for the relevant oscillation observables and provide a systematic parametrization suited to the study of non-standard effects in current and future large-baseline reactor neutrino data. As a concrete application, we perform a first analysis of the JUNO results using this approach, illustrating the sensitivity of the experiment to EFT effects.

The paper is organized as follows. In Sec.~\ref{sec:compact-formula} we present and develop the theoretical framework. Sec.~\ref{sec:EFT_ladder} provides a brief discussion of the EFT interactions relevant for reactor neutrino experiments. In Sec.~\ref{sec:JUNO-th} we derive the EFT-modified oscillation probability for a large-baseline reactor neutrino experiment, and Sec.~\ref{sec:pheno_at_JUNO} presents the application of our formalism to recent JUNO data. We conclude in Sec.~\ref{sec:conc}.

%%%%%%%%%%%%%%%%%%%%%%%%%%%%%%%%%%%%%%%%%%%%%%%%%%%%%%%%%%%%%%%%%%%%%%%%%%%%%%%%
\section{Theoretical Framework}
\label{sec:compact-formula}
%%%%%%%%%%%%%%%%%%%%%%%%%%%%%%%%%%%%%%%%%%%%%%%%%%%%%%%%%%%%%%%%%%%%%%%%%%%%%%%%

In this section we start by reviewing the QFT formalism introduced in Ref.~\cite{Falkowski:2019kfn} to describe neutrino oscillations in the presence of generic interactions. We then point out that the result can be conveniently written in a matrix form, which can be generalized in a natural way to include matter effects, which we discuss at the end of this section.

%%%%%%%%%%%%%%%%%%%%%%%%%%%%%%%%%%%%%%%%%%%%%%%%%%%%%%%%%%%%%%%%%%%%%%%%%%%%%%%%
\subsection{Oscillations in Vacuum}
\label{sec:vacuum}
%%%%%%%%%%%%%%%%%%%%%%%%%%%%%%%%%%%%%%%%%%%%%%%%%%%%%%%%%%%%%%%%%%%%%%%%%%%%%%%%
Let us consider a neutrino experiment where the neutrino is produced and detected through the CC processes $S \to X_\alpha \nu$ and $\nu\, T \to Y_\beta$, respectively, where $X_\alpha$ and $Y_\beta$ are (possibly multi-body) states containing the charged leptons $\alpha$ and $\beta$ ($\alpha,\beta = e,\mu,\tau$), the source $S$ and target $T$ are at rest and separated by a macroscopic distance $L$ in vacuum, and the neutrino emission is isotropic. 
The general QFT expression for the differential rate of detected events per target particle is given by~\cite{Falkowski:2019kfn}\footnote{See Refs.~\cite{Falkowski:2019kfn,Falkowski:2021bkq} for the generalization to multiple sources moving with different energies, multiple target particles, and non-isotropic neutrino emission.}
\begin{equation}\label{eq:event-rate-formula}
    R_{\alpha\beta} = \cfrac{dN^{\alpha\beta}}{N_T\,dt \,dE_\nu} =  \cfrac{N_S}{32\pi L^2 m_S m_T E_\nu}\sum_{k,\ell}e^{-i\frac{L\Delta m^2_{k\ell}}{2E_\nu}}\int d\Pi_{P^\prime} \cM^P_{\alpha k}{\bar\cM}^P_{\alpha \ell}\int d\Pi_{D} \cM^D_{\beta k}{\bar\cM}^D_{\beta \ell}\,.
\end{equation}
Here, $E_\nu$ is the (anti-)neutrino energy, $\Delta m_{k\ell}^{2}\equiv m_{k}^{2}-m_{\ell}^{2}$ is the mass squared difference between neutrino (mass) eigenstates (indexed by $k,\ell$), $\cM_{\alpha k}^P$  and $\cM_{\beta k}^D$ are the QFT amplitudes for the production and detection processes, and the bar indicates complex conjugation. The source and target masses are denoted by $m_{S}, m_{T}$, and $N_S, N_T$ are their corresponding number of particles. The phase space elements for production ($d \Pi_{P}$) and detection processes ($d \Pi_{D}$) are defined as usual, {$d \Pi \equiv \frac{d^3 k_1}{(2 \pi)^3 2 E_1} \dots  \frac{d^3 k_n}{(2 \pi)^3 2 E_n} (2\pi)^4 \delta^4(\sum p_j - \sum k_i )$}, with $k_i$ ($p_j$) and $E_i$ being the 4-momenta and energies of the final (initial) states. The primed phase space element for production is defined as $d \Pi_{P}\equiv d \Pi_{P^{\prime}} dE_{\nu} $. Sums and averages over spin and other unobserved degrees of freedom are implicitly included in the integral.

We now note that Eq.~\eqref{eq:event-rate-formula} can be recast into a compact matrix formula through a simple but enlightening rearrangement, namely
\begin{align}\label{eq:compact-rate}
    R_{\alpha\beta} 
    & = \Tr\left[ \cF \,\cfrac{d\Phi^\alpha} {dE_\nu}\,\cF^\dagger\,\Sigma^\beta\right]\,, 
\end{align}
where we define the following matrix quantities\footnote{Note that (i) the result is formally identical for neutrinos and antineutrinos; and (ii) generalized flux and cross-section matrices carry the mass-eigenstate indices in opposite order.}
\begin{align}
    \label{eq:define-flux-xsec}
    \frac{d\Phi^\alpha_{k\ell}}{dE_\nu} 
    &\equiv \frac{N_S}{8\pi L^2 m_S} \int d \Pi_{P^{\prime}} \mathcal{M}^{P}_{\alpha k} \bar{\mathcal{M}}^{P}_{\alpha \ell}\,,
    \\[3pt]
    \label{eq:define-xsec}
    \Sigma^\beta_{\ell k} 
    &\equiv \frac{1}{4m_T E_\nu} \int d \Pi_D \mathcal{M}^{D}_{\beta k} \bar{\mathcal{M}}^{D}_{\beta \ell}\,,
    \\[3pt]
    \label{eq:define-calF}
    \cF &\equiv e^{-i\,L\,{\cal H}} ~\to~ \cF_{k'k} = \delta_{k'k}\,\exp\left(-i\frac{L\,m_k^2}{2E_\nu}\right)\,.
\end{align}
We will refer to them as generalized (differential) flux, generalized cross section, and evolution matrix. This terminology follows Ref.~\cite{Coloma:2022umy}, adapted here to our QFT framework. In particular, because we do not assume SM production, the description involves not only a generalized cross section, but also a generalized flux. In Eqs.~\eqref{eq:define-flux-xsec}--\eqref{eq:define-calF} these quantities are given in the mass basis, i.e., in the basis where the free Hamiltonian takes the form ${\cal H}=M_d^2/(2E_\nu)$, where $M_d={\rm diag}(m_1^2,m_2^2,m_3^2)$ is the (diagonal) neutrino mass matrix. 
Let us note that both $k$ and $\ell$ subindices in the generalized flux $\Phi^\alpha$ refer to final neutrino (mass) states, in the processes $S\to X_\alpha \nu_k$ and $S\to X_\alpha \nu_\ell$. Thus, only the diagonal entries can be interpreted as ordinary fluxes, whereas the off-diagonal ones are associated to interference terms between diagrams mediated by different neutrino states. Likewise for the generalized cross section, where both subindices refer to the initial neutrino (mass) states~\cite{Coloma:2022umy}.

The intuitive picture behind this formula is quite simple: the total amplitude for the process can be calculated as the product of the amplitude for emitting $\nu_k$ (described by $\cM^P_{\alpha k}$), the probability of $\nu_k\to\nu_{k'}$ during propagation (described by $\cF_{k'k}$) and the amplitude for detecting $\nu_{k'}$ (described by $\cM^D_{\beta k'}$), leading to
\begin{align}\label{eq:sketch-mass}
A_{\alpha\to\beta\,}^{\sscript{tot}} &= ~\sum_{k,k'} ~\cM_{\alpha k}^P~ \times ~\cF_{k'k}~ \times ~ \cM_{\beta k'}^D~.
\end{align}
Taking the squared modulus of this amplitude and integrating over phase space we obtain  Eq.~\eqref{eq:compact-rate}, whereas Eq.~\eqref{eq:event-rate-formula} is recovered in vacuum, where $\cF_{k'k} = \delta_{k'k}\,\exp\left(-i\frac{L\,m_k^2}{2E_\nu}\right)$. That is, in the mass basis there is no oscillation between various neutrino states (because the evolution matrix $\cF$ is diagonal), but there is interference between diagrams mediated by different neutrino states (because the production/detection amplitudes are not diagonal).

The trace in the matrix expression for the rate given by Eq.~\eqref{eq:compact-rate} shows that the  result is basis independent. In the flavor basis we then have
\begin{align}\label{eq:mass-to-flavor}
    \Phi^\alpha_{\sscript{flavor}} &= U\,\Phi^\alpha_{\sscript{mass}}\,U^\dagger
    \quad \to\quad~\left[\frac{d\Phi^\alpha_{\sscript{flavor}}}{dE_\nu}\right]_{\eta\rho} = \frac{N_S}{8\pi L^2 m_S} \int d \Pi_{P^{\prime}} \mathcal{M}^{P}_{\alpha \eta} \bar{\mathcal{M}}^{P}_{\alpha \rho}\,,
    \\[3pt]
    \label{eq:mass-to-flavor2}\Sigma^\beta_{\sscript{flavor}} &= U\,\Sigma^\beta_{\sscript{mass}}\,U^\dagger
    \quad
    \to\quad~\lzs \Sigma^\beta_{\sscript{flavor}} \dzs_{\rho\eta} = \frac{1}{4m_T E_\nu} \int d \Pi_D \mathcal{M}^{D}_{\beta \eta} \bar{\mathcal{M}}^{D}_{\beta \rho}\,,
    \\[3pt]
    \label{eq:mass-to-flavor3}
    F_{\sscript{flavor}} &= U\,\cF_{\sscript{mass}} \,U^\dagger = e^{-i\,L\,H}\,.
\end{align}
In these expressions, $U$ denotes the PMNS matrix, $H=U\,{\cal H}\,U^\dagger$ is the Hamiltonian in the flavor basis, whereas $\cM^P_{\alpha \eta}=U_{\eta k}\cM^P_{\alpha k}$ and $\cM^D_{\beta \eta}=U^*_{\eta k}\cM^D_{\beta k}$ are the amplitudes for producing and detecting the flavor state $\nu_\eta$, respectively.\footnote{For antineutrinos, the flavor-to-mass basis transformation is $\cM^P_{\alpha \eta}=U^*_{\eta k}\cM^P_{\alpha k}$ and $\cM^D_{\beta \eta}=U_{\eta k}\cM^D_{\beta k}$.}
The intuitive picture in the flavor basis remains unchanged, namely
\begin{align}\label{eq:sketch-flavor}
A_{\alpha\to\beta\,}^{\rm tot} & = ~\sum_{\rho,\eta} ~\cM_{\alpha \rho}^P~ \times ~F_{\eta\rho}~ \times ~ \cM_{\beta \eta}^D~.
\end{align}
If SM and BSM interactions are diagonal in this basis, there is no lepton-flavor violation in production/detection and hence the corresponding amplitudes are proportional to $\delta_{\alpha\rho}$ and $\delta_{\beta\eta}$, whereas the $F$ matrix is not diagonal. In that case, one has oscillations between neutrino flavor states, while interference between amplitudes mediated by different neutrino states is absent. Note, however, that in general the three quantities in Eq.~\eqref{eq:sketch-flavor} are non-diagonal, giving rise to both flavor oscillations and interference effects. If Wilson Coefficients (NSIs) are defined in the flavor basis (as usually done), the generalized flux and cross section do not involve the $U$ matrix, which in this basis only appears in the evolution matrix $F$.

The notation introduced in this section makes particularly simple the generalization to the case of propagation in matter, which we discuss next.

%%%%%%%%%%%%%%%%%%%%%%%%%%%%%%%%%%%%%%%%%%%%%%%%%%%%%%%%%%%%%%%%%%%%%%%%%%%%%%%%
\subsection{Oscillations in Matter}
\label{sec:matter}
%%%%%%%%%%%%%%%%%%%%%%%%%%%%%%%%%%%%%%%%%%%%%%%%%%%%%%%%%%%%%%%%%%%%%%%%%%%%%%%%
As in the vacuum case, one can derive the formula for oscillations in non-uniform matter following a QFT approach where neutrino production, propagation and detection are considered as a single process. The approach followed in Ref.~\cite{Akhmedov:2012mk}, based partly on Ref.~\cite{Cardall:1999bz}, shows that the traditional QM approach is recovered for  ultrarelativistic neutrinos, with an effective potential that is (i) small compared with the neutrino energy, (ii) constant in time, and (iii) possibly non-constant in space, but with small variation compared to the localization regions of neutrino production and detection, which are assumed to be coherent. 

In addition to these general conditions, Ref.~\cite{Akhmedov:2012mk} assumes that neutrino interactions are purely SM-like. However, the derivation can be generalized to the case of generic low-energy contact interactions among SM fields, as these interactions can be recast as an effective local potential governing the neutrino evolution. In that case, the effective potential and the production and detection amplitudes are non-diagonal in flavor space, implying that the flavor basis cannot be defined unambiguously~\cite{Falkowski:2019kfn}.

This generalized derivation allows us to extend the framework presented in the previous subsection, for oscillations in vacuum, to the case of oscillations in a medium. We find that Eqs.~\eqref{eq:compact-rate}--\eqref{eq:define-calF} hold with only one simple and intuitive replacement: the Hamiltonian entering the evolution operator $\cF$ is now the full Hamiltonian ${\cal H}({\bf x})=\frac{M_d^2}{2E_\nu} + {\cal V}({\bf x})$, which includes the effect of the matter potential, determined by neutrino interactions with matter and by the matter density, which may in general depend on position. That is, the vacuum evolution operator $\cF=e^{-iL{\cal H}}$ is generalized to the path-ordered exponential $\cF=\mathcal P\exp\,[-i\int_C\,{\cal H}\,ds]$, where the Hamiltonian is integrated along the neutrino trajectory $C$, which connects the production and detection points (${\bf x}_0$ and ${\bf x}$, respectively). In other words, the evolution operator satisfies the Schr\"odinger-like equation $i\frac{d\cF}{dx}={\cal H}\,\cF$, with the boundary condition $\cF=\mathds{1}$ for $L\equiv|{\bf x}-{\bf x}_0|=0$. 
Here $\frac{d}{dx}\equiv{\hat{\bm r}}\cdot\!\boldsymbol{\nabla}$ denotes the directional derivative along the unit vector in the direction of the neutrino propagation: ${\hat{\bm r}}\equiv ({\bf x}-{\bf x}_0)/|{\bf x}-{\bf x}_0|$~\cite{Akhmedov:2012mk}.

As the result is basis independent, one can use instead the flavor basis, where the evolution operator satisfies the equation
\begin{align}
    i\frac{dF}{dx}=H\,F~,
\end{align} 
where 
\begin{align}
H({\bf x})
=\frac{M^\dagger M}{2E_\nu} + V({\bf x}) 
= \frac{UM_d^2 U^\dagger}{2E_\nu} + U\,{\cal V}({\bf x})\,U^\dagger~.
\end{align}
Thus, in this basis $F_{\beta\alpha}$ is nothing but the $\nu_\alpha\to\nu_\beta$ transition amplitude in the standard QM approach to neutrino oscillations in matter. In other words, the QFT derivation validates the traditional QM approach for the calculation of the evolution matrix, even in the presence of generic (heavy) BSM interactions of left-handed neutrinos. Let us emphasize the difference with the case of CC interactions, where the QFT approach reveals the limitations of the QM description based on source and detection NSIs, and makes it necessary to connect the latter to the underlying Lagrangian~\cite{Falkowski:2019kfn}.

Another interesting basis is the so-called matter basis, where the full Hamiltonian (including the matter potential) is diagonal, i.e., the basis where the Hamiltonian takes the form $\tilde{\cal H}=\tilde M_d^2/(2E_\nu)$, where $\tilde M_d={\rm diag}(\tilde m_1^2,\tilde m_2^2,\tilde m_3^2)$ is the (diagonal) effective neutrino mass matrix. In the general case of non-constant matter, this basis is position dependent and therefore differs between the production and detection points. It is then customary to present the result in an admixture of both bases, namely
 \begin{align}\label{eq:flavor-to-matter}
     \Phi^\alpha_{\sscript{matter}} &= \tilde U({\bf x}_0)^\dagger\,\Phi^\alpha_{\sscript{flavor}}\,\tilde U({\bf x}_0)~,
     \\[2pt]
     \Sigma^\beta_{\sscript{matter}} &= \tilde U({\bf x})^\dagger \,\Sigma^\beta_{\sscript{flavor}}\,\tilde U({\bf x})~,
     \\[2pt]
     \tilde \cF \,\equiv \cF_{\sscript{matter}}
     &= \tilde U({\bf x})^\dagger\,F(E_\nu,x,{\bf x}_0)_{\sscript{flavor}}\,\tilde U({\bf x}_0)~.
\end{align}
where $\tilde U({\bf x})$ and $\tilde U({\bf x}_0)$ are the rotation matrices that connect the flavor basis and the matter basis at the detection point $x$ and production point ${\bf x}_0$ respectively. 

In the adiabatic regime, $\tilde \cF$ is diagonal and its value has to be calculated numerically, except in some simple situations. 
For instance, in the case of constant density, the quantities $\tilde U$, $\tilde {\cal H}$ and $\tilde m_k$ are constant, and thus $\tilde \cF_{k'k} = \delta_{k'k}\,\exp\left(-i\frac{L\,\tilde m_k^2}{2E_\nu}\right)$. 
Thus, Eq.~\eqref{eq:event-rate-formula} holds with two simple changes: (i) neutrino masses must be replaced by the effective masses of the matter eigenstates in the medium: $m_k^2\to\tilde m_k^2$; and (ii) the production and detection amplitudes must be written in the matter basis, i.e., $\cM_{\alpha k}^P\to \cM_{\alpha \tilde k}^P \equiv \cM(S \to X_\alpha \nu_{\tilde k})$, where $\nu_{\tilde k}$ denotes the neutrino matter eigenstate, i.e., the mass eigenstate in the medium, indicated by a tilde on the index, and similarly for the detection amplitude. Taken together, we have
\begin{eqnarray}\label{eq:event-rate-formula-matter}
    R_{\alpha\beta}
    =  \cfrac{N_S}{32\pi L^2 m_S m_T E_\nu}\sum_{\tilde k,\tilde \ell}e^{-i\frac{L\Delta \tilde m^2_{\tilde k \tilde \ell}}{2E_\nu}}\int d\Pi_{P^\prime} \cM^P_{\alpha \tilde k}{\bar\cM}^P_{\alpha \tilde \ell}\int d\Pi_{D} \cM^D_{\beta \tilde k}{\bar\cM}^D_{\beta \tilde \ell}~.
\end{eqnarray}
Once the production and detection amplitudes are EFT decomposed, this gives the analytical expression presented in Ref.~\cite{Kopp:2025ffx}, where the case of non-constant density profiles is treated through the usual slab approximation, dividing the full baseline into segments of constant density.

%%%%%%%%%%%%%%%%%%%%%%%%%%%%%%%%%%%%%%%%%%%%%%%%%%%%%%%%%%%%%%%%%%%%%%%%%%%%%%%%
\subsection{Density Matrix and Oscillation Probability}
\label{sec:probability}
%%%%%%%%%%%%%%%%%%%%%%%%%%%%%%%%%%%%%%%%%%%%%%%%%%%%%%%%%%%%%%%%%%%%%%%%%%%%%%%%
Let us introduce the (conventional) neutrino flux and cross-section as
\begin{align}
\frac{d\phi^\alpha}{dE_\nu} &\equiv \Tr\, \frac{d\Phi^\alpha}{dE_\nu} = \frac{N_S}{8\pi L^2 m_S} \sum_\rho \int d \Pi_{P^{\prime}} |\mathcal{M}^{P}_{\alpha \rho}|^2~, 
\\ 
\sigma^\beta &\equiv ~{\rm Tr}\,\Sigma^\beta ~= \frac{1}{4m_T E_\nu} \sum_\rho \int d \Pi_D |\mathcal{M}^{D}_{\beta \rho}|^2~,
\end{align}
which are numbers (and not matrices in flavor space), unlike their generalized counterparts $\Phi^\alpha$ and $\Sigma^\beta$. While expressed above using the flavor-basis amplitudes, the flux and cross-section can be equivalently formulated with the mass-basis amplitudes owing to basis invariance. In a general basis, the quantity $\sigma^\beta$ is the sum of three cross-sections, associated with the processes $\nu_\rho\,T\to\,Y_\beta$, where $\rho$ runs over the three neutrino states. In the SM case, only the $\rho=\beta$ state yields a non-zero cross-section (independently of the detection process); thus $\sigma^\beta$ reduces to the usual SM cross-section. An identical argument applies to the flux.

Using the flux introduced above, the event rate in Eq.~\eqref{eq:compact-rate} can be rewritten in a more conventional form
\begin{align}\label{eq:compact-rate_with_rho}
    R_{\alpha\beta} 
    & = \cfrac{d\phi^\alpha} {dE_\nu}\,\Tr\left[ \rho \,\Sigma^\beta\right]\,, 
\end{align}
where the QFT expression for the density matrix is given by
\begin{align}\label{eq:rho_def}
    \rho=F \rho_0\, F^\dagger~,
    \quad\quad \rho_0=\frac{d\Phi^\alpha}{dE_\nu}\lzm \frac{d\phi^\alpha}{dE_\nu} \dzm^{\eminus1}~.
\end{align}
For notational simplicity, we suppress the superscript $\alpha$ in $\rho$. We show that this quantity satisfies the standard properties of a quantum density matrix in App.~\ref{app:rho}.

The density matrix $\rho$ encodes the quantum state of the neutrino after propagation over a distance $L$ from the production point. Equivalently, the quantity $\phi^\alpha \rho$ corresponds to the generalized flux evaluated at the detection point. We emphasize that, in general, $\rho$ depends on the production process, which may involve strong, weak, and possible non-standard interactions. However, for SM production one has $\Phi^\alpha = \phi^\alpha_{\sscript{SM}}\, \Pi^\alpha$, where $\Pi^\alpha$ denotes the flavor projection matrix, whose only nonzero entry is $[\Pi^\alpha]_{\alpha\alpha}=1$, i.e. $[\Pi^\alpha]_{\rho\eta} = \delta_{\rho\alpha}\delta_{\alpha\eta}$. Thus, we have $\rho=F\,\Pi^\alpha\, F^\dagger$, and the dependence on the production process disappears, apart from the flavor projector itself.

Combining the generalized flux and cross section, the oscillation probability $P_{\alpha\beta}$ can be defined by~\cite{Falkowski:2019kfn}
\begin{equation}
\label{eq:probability-def}
R_{\alpha\beta} = \frac{d\phi^\alpha}{dE_\nu} \, P_{\alpha\beta}\,\sigma^\beta~,
\end{equation}
which, using the QFT expression for the rate in Eq.~\eqref{eq:compact-rate}, gives
\begin{equation}
\label{eq:probability}
P_{\alpha\beta}
= \frac{\Tr\left[ F\,\cfrac{d\Phi^\alpha} {dE_\nu}\,F^\dagger\,\Sigma^\beta \right]}{\Tr\left[ \cfrac{d\Phi^\alpha} {dE_\nu}\right] \Tr\left[ \Sigma^\beta \right]}
~=~ \frac{\Tr\left[ \rho\,\Sigma^\beta \right]}{\Tr\left[ \Sigma^\beta \right]}~,
\end{equation}
which can be shown to be bounded between zero and unity (see App.~\ref{sec:proof}), hence the name oscillation {\it probability}.\footnote{One should note, however, that the unitarity constraint, $\sum\limits_\alpha P_{\alpha\beta}=\sum\limits_\beta P_{\alpha\beta}=1$, is not satisfied in general.} 

In general, the generalized flux and cross-section do not cancel between numerator and denominator of Eq.~\eqref{eq:probability}. Consequently, the oscillation probability depends on the specific production and detection processes and is, therefore, not universal. Equivalently, the process lacks a strict factorization into production, oscillation, and detection stages. While the QFT derivation in Ref.~\cite{Akhmedov:2012mk} demonstrates that factorization holds under SM interactions, this property does not extend to our framework, which incorporates general BSM interactions. This departure from factorization is expected, as it is known to occur even in vacuum under similar conditions~\cite{Falkowski:2019kfn}. In the SM case, one has $\Phi^\alpha = \phi^\alpha_{\sscript{SM}} \Pi^\alpha$ (and similarly for the cross section), so that production and detection effects cancel in the probability, yielding the well-known result $P_{\alpha\beta}=\tr(F\Pi^\alpha F^\dagger \Pi^\beta)=|F_{\beta\alpha}|^2$.

Finally, one can also define an {\it effective} oscillation probability (or {\it pseudo}-probability) by
\begin{align}
\label{eq:P_eff}
    R_{\alpha\beta} = \frac{d\phi^\alpha_{\sscript{SM}}}{dE_\nu} \,\widetilde P_{\alpha\beta}\,\sigma^\beta_{\sscript{SM}}\,.
\end{align}
Comparing with Eq.~\eqref{eq:probability-def} one trivially sees that
\begin{align}
   \widetilde P_{\alpha\beta} = 
   P_{\alpha\beta} \,
   \lzm \frac{d\phi^\alpha}{dE_\nu}\,\sigma^\beta \dzm \lzm \frac{d\phi^\alpha_{\sscript{SM}}}{dE_\nu}\,\sigma^\beta_{\sscript{SM}} \dzm^{\eminus1}~,
\end{align}
which shows that the effective probability is not between 0 and 1 unless production and detection are free from BSM effects. 
Nonetheless, it can be convenient to use the effective probability in some cases, as it encodes all BSM effects affecting the rate. It is also a useful quantity for many phenomenological analyses, since SM production and detection is typically assumed in the experimental extraction of the probability. That is, the quantity extracted in experimental analyses often corresponds to this effective probability.

It is important to note that the calculation of the SM fluxes and cross-section depends typically on certain parameters, like $V_{ud}$ or $g_A$, which are extracted from other experiments under the assumption of the SM. This introduces additional {\it indirect} BSM effects, which are in general of the same order as the regular {\it direct} BSM effects~\cite{Falkowski:2019xoe}. If the flux-cross-section product is not calculated but measured using a near detector, one should consider that the (actual or effective) transition probability is, in general, not one at zero distance~\cite{Falkowski:2019kfn}.

%%%%%%%%%%%%%%%%%%%%%%%%%%%%%%%%%%%%%%%%%%%%%%%%%%%%%%%%%%%%%%%%%%%%%%%%%%%%%%%%
\subsection{NC detection}
%%%%%%%%%%%%%%%%%%%%%%%%%%%%%%%%%%%%%%%%%%%%%%%%%%%%%%%%%%%%%%%%%%%%%%%%%%%%%%%%
Our results can be easily extended to the case of NC detection, $\nu_k T\to \nu_j Y$.  
The only differences with respect to the CC detection are the following: (i) we sum over the final state index $j$, as we have no information about the neutrino final mass eigenstate (or flavor); and (ii) we work with the differential cross-section with respect to a generic kinematic variable $T$ that might be accessible experimentally and thus it should not be integrated over. For instance, in coherent elastic neutrino–nucleus scattering (CE$\nu$NS), $T$ corresponds to the recoil energy, $T=E_T' - m_T$.

The QFT expression for the rate was presented in Ref.~\cite{Breso-Pla:2023tnz} for the case of propagation in vacuum. Proceeding as in the case of CC detection, we can express it in matrix form and generalize it for propagation in matter:
\begin{equation}
\label{eq:DifferentialNumberEventsFunctionR}
R_{\alpha} (t,T,E_\nu)
\equiv \frac{1}{N_T}\frac{dN^{\alpha}}{dt \, dE_{\nu}\, dT}
= {\rm Tr} \lzs F \,\frac{d\Phi^\alpha}{dE_\nu} \,F^\dagger \,\frac{d\Sigma}{dT} \dzs
= \frac{d\phi^\alpha}{dE_\nu}\, {\rm Tr} \lzs \rho \,\frac{d\Sigma}{dT} \dzs~.
\end{equation}
Here the differential cross-section is given by 
    \begin{align}
    \frac{d\Sigma_{\eta\rho}}{dT} &\equiv \frac{1}{4m_T E_\nu} \sum_\gamma \int d \Pi_{D^{\prime}} \mathcal{M}^{D}_{\gamma \rho} \bar{\mathcal{M}}^{D}_{\gamma \eta}~,
    \end{align}
where $\mathcal{M}^{D}_{\gamma \rho}={\cal M}\left(  \nu_\rho T \to \nu_\gamma Y  \right)$, and $d \Pi_{D}\equiv d \Pi_{D^{\prime}} dT $. The sum over $\gamma$ implies that $\Sigma=\Sigma^e+\Sigma^\mu+\Sigma^\tau$, where $\Sigma^\gamma$ is defined as in the CC case with the obvious replacement of a charged lepton $\ell_\gamma$ by a neutrino $\nu_\gamma$ in the final state. We discuss some interesting limits:
\begin{itemize}
\item As discussed before, one has $\frac{d\Phi^\alpha}{dE_\nu} = \frac{d\phi^\alpha_{\sscript{SM}}}{dE_\nu} \Pi^\alpha$ in the case of SM production, and hence $\rho=F\Pi^\alpha F^\dagger$, i.e., the density matrix becomes independent of the production details. Thus we recover Eq.~(2.4) in Ref.~\cite{Coloma:2022umy}.
\item If the cross-section is flavor conserving and universal, we have $\frac{d\Sigma}{dT}=\frac{d\sigma}{dT}\,\mathbb{1}$. Then $F$ and $F^\dagger$ cancel (using the cyclic property of the trace), i.e., the rate becomes independent of the oscillation parameters. In particular, we have $R_\alpha=\frac{d\phi^\alpha}{dE_\nu}\,\frac{d\sigma}{dT}$. This happens, e.g., in the SM for CE$\nu$NS detection, but not for electron neutrino elastic scattering.
\end{itemize}

%%%%%%%%%%%%%%%%%%%%%%%%%%%%%%%%%%%%%%%%%%%%%%%%%%%%%%%%%%%%%%%%%%%%%%%%%%%%%%%%
\section{The EFT Ladder}
\label{sec:EFT_ladder}
%%%%%%%%%%%%%%%%%%%%%%%%%%%%%%%%%%%%%%%%%%%%%%%%%%%%%%%%%%%%%%%%%%%%%%%%%%%%%%%%
In Sec.~\ref{sec:compact-formula} we outlined a general QFT description of neutrino oscillation experiments and obtained a compact expression for the event rate that systematically incorporates possible NP effects in neutrino production, propagation and detection. To describe such effects in a model-independent way, one can adopt an EFT-ladder approach, connecting the relevant effective descriptions at different energy scales, which provides a convenient parametrization of non-standard interactions (NSI). In the following, we focus on CC semileptonic interactions, as these are the relevant ones for our subsequent analysis of medium-baseline reactor experiments. This EFT ladder was discussed in Ref.~\cite{Falkowski:2019xoe}, and here we summarize its main elements for self-consistency.

%%%%%%%%%%%%%%%%%%%%%%%%%%%%%%%%%%%%%%%%%%%%%%%%%%%%%%%%%%%%%%%%%%%%%%%%%%%%%%%%
\subsection{From SMEFT to LEFT}
%%%%%%%%%%%%%%%%%%%%%%%%%%%%%%%%%%%%%%%%%%%%%%%%%%%%%%%%%%%%%%%%%%%%%%%%%%%%%%%%
At energies above the electroweak scale, the SMEFT offers a model-independent description of heavy new physics, whose effects are encoded in gauge-invariant higher-dimensional operators constructed from SM fields. Within this framework, the first correction to the SM is the dimension-5 Weinberg operator~\cite{Weinberg:1979sa}, which generates Majorana masses for the neutrinos. Their interactions are modified through dimension-six operators. For reactor neutrino experiments in which matter effects can be neglected, the relevant operators are those that modify CC semileptonic interactions, either through corrections to the SM gauge-boson couplings or new four-fermion contact interactions involving leptons and quarks.

%%%%%%%%%%%%%%%%%%%%%%%%%%%%%%%%%%%%%%%%%%%%%%%%%%%%%%%%%%%%%%%%%
%%%%%%%%%%%%%%%%%%%%%%%%%%%%%%%%%%%%%%%%%%%%%%%%%%%%%%%%%%%%%%%%%
\begin{table}[t]
\centering
\scalebox{0.9}{
\begin{tabular}{c@{\hspace{1.0cm}}c@{\hspace{1.0cm}}c@{\hspace{1.0cm}}c}
\toprule
\textbf{Operator}
&\textbf{Definition}
&\textbf{Operator}
&\textbf{Definition}
\\
\midrule
\addlinespace[0.2cm]
%%%%%%%%%%%%%%%%%%%%%%%%%%%%%%%%%%%%%%%%%
$[\cO_{H\ell}^{(3)}]_{\alpha\beta}$
&$(H^\dag i\overset{\text{\footnotesize$\leftrightarrow$}}{D^a_\mu} H)(\bar\ell_\alpha\gamma^\mu\sig^a\ell_\beta)$
&$[\cO_{\ell q}^{(3)}]_{\alpha\beta ij}$
&$(\bar\ell_{\alpha}\gamma^\mu\sigma^a \ell_{\beta})(\bar q_{i}\gamma_\mu \sigma^a q_{j})$
%%%%%%%%%%%%%%%%%%%%%%%%%%%%%%%%%%%%%%%%%
\\
\noalign{\vskip 0.2cm}
% \cmidrule{2-4}
\cdashline{1-4}[.4pt/2pt]
\noalign{\vskip 0.2cm}
%%%%%%%%%%%%%%%%%%%%%%%%%%%%%%%%%%%%%%%%%
$[\cO_{Hq}^{(3)}]_{ij}$
&$(H^\dag i\overset{\text{\footnotesize$\leftrightarrow$}}{D^a_\mu} H)(\bar q_i\gamma^\mu\sig^a q_j)$
&$[\cO_{\ell equ}^{(1)}]_{\alpha\beta ij}$
&$(\bar\ell_\alpha e_\beta)i\sigma_2(\bar q_i u_j)^\intercal$
%%%%%%%%%%%%%%%%%%%%%%%%%%%%%%%%%%%%%%%%%
\\
\noalign{\vskip 0.2cm}
% \cmidrule{2-4}
\cdashline{1-4}[.4pt/2pt]
\noalign{\vskip 0.2cm}
%%%%%%%%%%%%%%%%%%%%%%%%%%%%%%%%%%%%%%%%%
$[\cO_{Hud}]_{ij}$
&$(\widetilde H^\dag iD_\mu H)(\bar u_{i}\gamma^\mu d_{j})$
&$[\cO_{\ell equ}^{(3)}]_{\alpha\beta ij}$
&$(\bar\ell_\alpha \sigma^{\mu\nu} e_\beta)i\sigma_2(\bar q_i \sigma_{\mu\nu} u_j)^\intercal$
%%%%%%%%%%%%%%%%%%%%%%%%%%%%%%%%%%%%%%%%%
\\
\noalign{\vskip 0.2cm}
% \cmidrule{2-4}
\cdashline{1-4}[.4pt/2pt]
\noalign{\vskip 0.2cm}
%%%%%%%%%%%%%%%%%%%%%%%%%%%%%%%%%%%%%%%%%
&&
$[\cO_{\ell edq}]_{\alpha\beta ij}$
&$(\bar\ell_\alpha e_\beta)(\bar d_i q_j)$
\\[0.2cm]
\bottomrule
\end{tabular}
}
\caption{Overview of the relevant SMEFT operators generating tree-level NSIs in Eq.~\eqref{eq:eps_SMEFT_LEFT_matching}.}
\label{tab:SMEFT_ops}
\end{table}
%%%%%%%%%%%%%%%%%%%%%%%%%%%%%%%%%%%%%%%%%%%%%%%%%%%%%%%%%%%%%%%%%
%%%%%%%%%%%%%%%%%%%%%%%%%%%%%%%%%%%%%%%%%%%%%%%%%%%%%%%%%%%%%%%%%

Below the electroweak scale, the heavy electroweak degrees of freedom are integrated out, and the appropriate description is provided by a low-energy effective field theory (LEFT) written in terms of light quarks, leptons, gluons, and photons. The effects of the SMEFT operators are matched onto local four-fermion operators governing CC processes. The leading LEFT CC semileptonic interactions can be written as\footnote{Note that both SMEFT and LEFT assume the absence of light right-handed neutrino states.}
\begin{equation}\label{eq:LEFT_Lagr}
    \begin{alignedat}{2}
        \cL_\sscript{LEFT}&\supset-\frac{2V_{ud}}{v^2}\Big\{ 
        [\mathds{1}+\epsilon_L]_{\alpha\beta}(\bar u \gamma^\mu P_L d)(\bar\ell_\alpha \gamma_\mu P_L \nu_\beta)
        +[\epsilon_R]_{\alpha\beta}(\bar u \gamma^\mu P_R d)(\bar\ell_\alpha \gamma_\mu P_L \nu_\beta)
        \\[0.1cm]
        &~~~~~~~~~~~~
        +\frac{1}{2}[\epsilon_S]_{\alpha\beta}(\bar ud)(\bar\ell_\alpha P_L\nu_\beta)
        -\frac{1}{2}[\epsilon_P]_{\alpha\beta}(\bar u\gamma_5 d)(\bar\ell_\alpha P_L\nu_\beta)
        \\[0.1cm]
        &~~~~~~~~~~~~
        +\frac{1}{4}[\epsilon_T]_{\alpha\beta}(\bar u \sigma^{\mu\nu}P_L d)(\bar\ell_\alpha \sigma_{\mu\nu}P_L \nu_\beta)+\hermc\Big\}\,.
    \end{alignedat} 
\end{equation}
Here $v\equiv(\sqrt2G_F)^{\eminus1/2}\approx246\,\gev$ denotes the Higgs vacuum expectation value, $V_{ud}$ is the CKM matrix element, $\ell_\alpha=e,\mu,\tau$ are charged-lepton fields, $\nu_\alpha$ are neutrino flavor eigenstates, $\sigma^{\mu\nu}=i[\gamma^\mu,\gamma^\nu]/2$, and $P_{L,R}=(1\mp\gamma_5)/2$. The coefficients $\epsilon_X$ are $3\times3$ matrices in lepton flavor space that encode non-standard interactions, with $X\in\{L,R,S,P,T\}$ labeling the different Lorentz structures. They are obtained by matching the LEFT onto the dimension-six SMEFT operators in the Warsaw basis~\cite{Grzadkowski:2010es} at the electroweak scale, where both corrections to the SM charged-current vertices and semileptonic contact interactions are generated~\cite{Cirigliano:2009wk,Cirigliano:2012ab,Aebischer:2017gaw,Gonzalez-Alonso:2017iyc,Falkowski:2019xoe}:
\begin{equation}\label{eq:eps_SMEFT_LEFT_matching}
    \begin{alignedat}{2}
        [\epsilon_L]_{\alpha\beta}&\approx \frac{1}{V_{ud}}\frac{v^2}{\Lambda^2}\lzm V_{ud}[\cC_{H\ell}^{(3)}]_{\alpha\beta}+V_{jd}[\cC_{Hq}^{(3)}]_{1j}\delta_{\alpha\beta}-V_{jd}[\cC_{\ell q}^{(3)}]_{\alpha\beta 1j} \dzm\,,
        \\[2pt]
        [\epsilon_R]_{\alpha\beta}&\approx \frac{1}{2V_{ud}}\frac{v^2}{\Lambda^2}[\cC_{Hud}]_{11}\delta_{\alpha\beta}\,,
        \\[2pt]
        [\epsilon_{S,P}]_{\alpha\beta}&\approx-\frac{1}{2V_{ud}}\frac{v^2}{\Lambda^2}\lzm V_{jd}[\cC_{\ell equ}^{(1)}]^*_{\beta\alpha j1}\,\pm\,[\cC_{\ell edq}]^*_{\beta\alpha 11} \dzm\,,
        \\[2pt]
        [\epsilon_T]_{\alpha\beta}&\approx -\frac{2}{V_{ud}}\frac{v^2}{\Lambda^2}V_{jd}[\cC_{\ell equ}^{(3)}]^*_{\beta\alpha j1}\,.
    \end{alignedat}
\end{equation}
Here, the Wilson coefficients on the right-hand side correspond to the operators listed in Tab.~\ref{tab:SMEFT_ops}, with the normalization ${\cal L}=\sum_i \frac{{\cal C}_i}{\Lambda^2}{\cal O}_i$.

Neutrino flavor eigenstates $\nu_\alpha$ are related to mass eigenstates $\nu_i$ through $\nu_\alpha = U_{\alpha i}\nu_i$, where $U$ denotes the PMNS matrix. In the SMEFT, since neutrinos and left-handed charged leptons form an $SU(2)_L$ doublet, $U$ arises as the relative rotation between their respective diagonalizing transformations in flavor space. In general, it does not necessarily render neutrino interactions flavor diagonal, although this is the case in specific scenarios such as the SM. The PMNS matrix is conventionally parametrized in terms of three mixing angles and a CP-violating phase $\delta_\sscript{CP}$
\begin{equation}\label{eq:PMNS-Matrix}
U =
\begin{bmatrix}
c_{12} c_{13}
& ~~s_{12} c_{13}
& ~~e^{\eminus i\delta_{\sscript{CP}}} s_{13}
\\[0.3em]
- s_{12} c_{23} - e^{i\delta_{\sscript{CP}}} c_{12} s_{13} s_{23}
& ~~\phantom{-}c_{12} c_{23} - e^{i\delta_{\sscript{CP}}} s_{12} s_{13} s_{23}
& ~~c_{13} s_{23}
\\[0.3em]
\phantom{-}s_{12} s_{23} - e^{i\delta_{\sscript{CP}}} c_{12} s_{13} c_{23}
& ~~- c_{12} s_{23} - e^{i\delta_{\sscript{CP}}} s_{12} s_{13} c_{23}
& ~~c_{13} c_{23}
\end{bmatrix}
\,,
\end{equation}
where $s_{ij}\equiv\sin\theta_{ij}$ and $c_{ij}\equiv\cos\theta_{ij}$.

%%%%%%%%%%%%%%%%%%%%%%%%%%%%%%%%%%%%%%%%%%%%%%%%%%%%%%%%%%%%%%%%%%%%%%%%%%%%%%%%
\subsection{Lee--Yang EFT}
\label{sec:LY_lagr_EFT}
%%%%%%%%%%%%%%%%%%%%%%%%%%%%%%%%%%%%%%%%%%%%%%%%%%%%%%%%%%%%%%%%%%%%%%%%%%%%%%%%
At the energies relevant for reactor neutrino experiments, it is pertinent to match the LEFT onto a low-energy description in terms of proton and neutron fields interacting with leptons and neutrinos. This procedure leads to an effective CC Lagrangian, commonly referred to as the Lee--Yang EFT~\cite{Lee:1956qn}, in which the short-distance physics encoded in the LEFT coefficients is absorbed into nucleon-level couplings multiplying the allowed Lorentz structures~\cite{Gonzalez-Alonso:2018omy}:
\begin{align}
        \cL_\sscript{LY}&\supset -\frac{V_{ud}}{v^2}
        \Big\{ 
        g_V[\mathds{1}+\epsilon_L+\epsilon_R]_{\alpha\beta}(\bar p\gamma^\mu n)(\bar\ell_\alpha \gamma_\mu P_L \nu_\beta) +g_S[\epsilon_S]_{\alpha\beta}(\bar p n)(\bar\ell_\alpha P_L \nu_\beta)
        \nonumber\\[3pt]
        &~~~~~~~~~~-g_A[\mathds{1}+\epsilon_L-\epsilon_R]_{\alpha\beta}(\bar p \gamma^\mu \gamma_5 n)(\bar\ell_\alpha \gamma_\mu P_L \nu_\beta)
        -g_P[\epsilon_P]_{\alpha\beta}(\bar p \gamma_5 n)(\bar\ell_\alpha P_L \nu_\beta)
        \nonumber\\[3pt]
        &~~~~~~~~~~+\frac{g_T}{2}[\epsilon_T]_{\alpha\beta}(\bar p\sigma^{\mu\nu}P_Ln)(\bar\ell_\alpha\sigma_{\mu\nu}P_L\nu_\beta) +\hermc\Big\}\,.
\end{align}
Here $p$ and $n$ denote the relativistic proton and neutron fields. The coefficients $g_X$, with $X\in\{V,A,S,P,T\}$, are the nucleon charges associated with the corresponding quark bilinears and arise from matching the quark-level operators in the LEFT onto nucleon matrix elements of the form $\bra{p}\bar u\,\Gamma_X d\ket{n}$. These quantities encode non-perturbative QCD effects and can be determined from lattice QCD or experimental input. In particular, conservation of the vector current implies $g_V=1$ up to small isospin-breaking corrections~\cite{Ademollo:1964sr}. The axial charge $g_A$ is extracted precisely from neutron beta decay, while the scalar and tensor charges $g_S$ and $g_T$ are typically taken from lattice calculations.\footnote{The phenomenological $g_A$ value is subject to corrections from new physics. However, as we employ it only for the calculation of NP contributions, such effects enter at next-to-leading order and are therefore negligible within our power counting. For analyses incorporating higher-order corrections, one may replace this value with lattice results; regardless, the numerical impact on our current results remains marginal.} The pseudoscalar charge $g_P$, which can be related to $g_A$, is strongly enhanced at low momentum transfer~\cite{Gonzalez-Alonso:2013ura}.
At the reference scale $\mu \simeq 2\,\gev$, the numerical values used in this work are $g_A\approx1.2728$, $g_S\approx1.02$, $g_P\approx349$ and $g_T\approx0.987$~\cite{Gonzalez-Alonso:2018omy,Bhattacharya:2016zcn,Gonzalez-Alonso:2013ura}. The corresponding uncertainties are sufficiently small to have a negligible impact on the results presented below and are therefore neglected throughout our analysis.

%%%%%%%%%%%%%%%%%%%%%%%%%%%%%%%%%%%%%%%%%%%%%%%%%%%%%%%%%%%%%%%%%%%%%%%%%%%%%%%%
\subsection{Non-Relativistic Limit of Lee--Yang EFT} 
%%%%%%%%%%%%%%%%%%%%%%%%%%%%%%%%%%%%%%%%%%%%%%%%%%%%%%%%%%%%%%%%%%%%%%%%%%%%%%%%
For reactor neutrino experiments, an additional simplification arises from the kinematics of the relevant processes. While the charged leptons and neutrinos are relativistic, the momentum transfer to the hadronic system is small compared to the nucleon mass. As a result, the proton and neutron fields can be treated in the non-relativistic limit, leading to a further reduction of the effective description. In this regime, the Lee--Yang Lagrangian can be expanded in powers of the momentum transfer over the nucleon mass, and the leading contributions are captured by a non-relativistic effective theory for the hadronic currents:
\begin{equation}\label{eq:LY_NR}
\begin{alignedat}{2}
\cL_\sscript{LY}^\sscript{NR}\supset 
&-\frac{V_{ud}}{v^2}(\psi_p^{\dagger}\psi_n)\Big\{[\mathds{1}+\epsilon_L+\epsilon_R]_{\alpha\beta}(\bar\ell_\alpha \gamma^0P_L \nu_\beta)+g_S[\epsilon_S]_{\alpha\beta}(\bar\ell_\alpha P_L \nu_\beta)\Big\}
\\[0.13cm]&
+\frac{V_{ud}}{v^2}(\psi_p^{\dagger}\sig^k\psi_n)\Big\{g_A[\mathds{1}+\epsilon_L-\epsilon_R]_{\alpha\beta}(\bar\ell_\alpha \gamma^kP_L \nu_\beta)-g_T[\epsilon_T]_{\alpha\beta}(\bar\ell_\alpha \gamma^0 \gamma^k P_L\nu_\beta) \Big\}
+\hermc\,,
\end{alignedat}
\end{equation}
where $\psi_p$ and $\psi_n$ denote the non-relativistic proton and neutron fields, respectively, and $\sigma^k$ are the Pauli matrices. The lepton bilinears, however, are not subject to a non-relativistic reduction, and we therefore retain the relativistic structures involving $\gamma^k$, matching the notation adopted in Ref.~\cite{Falkowski:2021vdg}.

The structure of Eq.~\eqref{eq:LY_NR} implies two immediate observations. First, pseudoscalar interactions do not appear at leading order in the non-relativistic expansion~\cite{Falkowski:2021vdg}. Second, the hadronic sector is fully described at leading order by two independent transition operators: the spin-independent $\psi_p^{\dagger}\psi_n$, corresponding to Fermi transitions, and the spin-dependent $\psi_p^{\dagger}\sigma^k\psi_n$, corresponding to Gamow--Teller transitions. They determine the leading nuclear response governing (inverse) beta decay in reactor neutrino experiments.

%%%%%%%%%%%%%%%%%%%%%%%%%%%%%%%%%%%%%%%%%%%%%%%%%%%%%%%%%%%%%%%%%%%%%
\section{New Physics at Reactor Neutrino Experiments}
\label{sec:JUNO-th}
%%%%%%%%%%%%%%%%%%%%%%%%%%%%%%%%%%%%%%%%%%%%%%%%%%%%%%%%%%%%%%%%%%%%%
In this section, we present a detailed discussion of the impact of New Physics on neutrino production and detection processes at reactor experiments~\cite{Falkowski:2019xoe}. We further describe how the effective transition and survival probabilities are computed within our framework, focusing in particular on medium-baseline facilities ($L \sim 50$ km).

%%%%%%%%%%%%%%%%%%%%%%%%%%%%%%%%%%%%%%%%%%%%%%%%%%%%%%%%%%%%%%%%%%%%%
\subsection{Production: Beta Decay}
%%%%%%%%%%%%%%%%%%%%%%%%%%%%%%%%%%%%%%%%%%%%%%%%%%%%%%%%%%%%%%%%%%%%%

At nuclear reactor facilities, antineutrinos are generated through beta decays involving a large variety of parent nuclei~\cite{Hayes:2016qnu}. The corresponding production amplitude $\cM^P_{\alpha \rho} \equiv \cM\left({\cN} \rightarrow \ell^{-}_\alpha\,\Bar{\nu}_{\rho}\,\cN^\prime\right)$ can be decomposed into SM and BSM contributions as
\begin{equation}\label{eq:prod-ampl}
    \cM^P_{\alpha \rho}= \delta_{\alpha\rho}\,A^P_L + \sum_{X}\, [\epsilon_{X}]_{\alpha\rho}\, A^P_X\,,
\end{equation}
where the dependence of the amplitudes $A_X^P$ on the charged-lepton flavor index $\alpha$ is suppressed for notational simplicity~\cite{Falkowski:2019kfn}. Starting from the Lagrangian in Eq.~\eqref{eq:LY_NR}, the amplitudes can be written as
\begin{eqnarray}\label{eq:A-X-P}
    A^P_{L/R} &=& -\cfrac{V_{ud}}{v^2} \left[\braket{\psi_p^{\dagger}\,\psi_n} (\bar{u}_{\ell_\alpha}\gamma^0 P_L v_{\nu_\rho} ) \,\mp\, g_A \braket{\psi_p^{\dagger}\,\sigma^k\,\psi_n} ( \bar{u}_{\ell_\alpha}\gamma^k P_L v_{\nu_\rho} ) \right] \,, 
    \nonumber\\[3pt] 
    A^P_S &=& -\cfrac{V_{ud}}{v^2}\, g_S \braket{\psi_p^{\dagger}\,\psi_n} ( \bar{u}_{\ell_\alpha}P_L v_{\nu_\rho} )\,,\qquad\qquad A^P_P = 0 \,,
    \nonumber\\[3pt]
    A^P_T &=& -\cfrac{V_{ud}}{v^2}\, g_T \braket{\psi_p^{\dagger}\,\sigma^k\,\psi_n} ( \bar{u}_{\ell_\alpha}\gamma^0\gamma^k P_L v_{\nu_\rho} )\,,
\end{eqnarray}
where the lepton fields are replaced by the corresponding Dirac spinors, and $\braket{\,\cdot\,}$ is shorthand for the matrix elements of the nucleons with respect to the parent and daughter nuclei states, i.e. $\braket{\,\cdot\,} \equiv \bra{\cN^\prime({\bm p}_{\cN^\prime})}\,\cdot\,\ket{\cN({\bm p}_{\cN})} $. Here, ${\bm p}_{\cN}$ (${\bm p}_{\cN^\prime}$) denotes the momentum of the parent (daughter) nucleus. These states satisfy the relativistic normalization $\braket{\cN^\prime({\bm p}_{\cN^\prime}) | \cN({\bm p}_{\cN})} = 2E_{\cN}(2\pi)^3\delta^3({\bm p}_{\cN} - {\bm p}_{\cN^\prime})$.

The general form of the nuclear matrix elements $\braket{\psi_p^{\dagger}\,\psi_n}$ and $\braket{\psi_p^{\dagger}\,\sigma^k\,\psi_n}$, based on rotational and Galilean symmetry, parity, time-reversal, and isospin symmetry, is~\cite{Falkowski:2021vdg}
\begin{equation}\label{eq:nuclear-matrix-elems}
\braket{\psi_p^{\dagger}\,\psi_n} =
2\,\sqrt{E_\cN\,E_{\cN^\prime}}\, M_F \, \delta_{J_z^\prime J_z}\,, 
\qquad
\braket{\psi_p^{\dagger}\,\sigma^k\,\psi_n} =
2\,\sqrt{E_\cN\,E_{\cN^\prime}}\,M_F\,\frac{r\,[{\cal T}^k_J]^{J_z}_{J_z^\prime}}{\sqrt{J(J+1)}}\,,
\end{equation}
where ${\cal T}^k_J$ are the spin-$J$ generators of the $\mathrm{SO}(3)$ Lie Algebra, $J_z$, $J_z^\prime$ the $z$-components of the total angular momenta of the two nuclei, $r$ parametrizes the ratio of the Fermi and Gamow--Teller matrix elements, and $M_F$ is a normalization factor characterizing the specific $\cN \to \cN^\prime$ transition. For $\beta^\mp$ decays between members of the same isospin multiplet, it reads
\begin{eqnarray}\label{eq:M-F}
    M_F = \delta_{j^\prime j}\delta_{j_3^\prime, j_3\pm1}\sqrt{j(j+1) - j_3(j_3 \pm 1)}\,,
\end{eqnarray}
with $(j, j_3)$ and $(j^\prime, j^\prime_3)$ being the isospin quantum numbers of $\cN$ and $\cN^\prime$ respectively.\footnote{For $\beta^+$ decays, Eqs.~\eqref{eq:A-X-P}--\eqref{eq:nuclear-matrix-elems} hold upon exchanging proton and neutron fields, $\braket{\psi_p^\dagger \psi_n} \to \braket{\psi_n^\dagger \psi_p}$, and making the corresponding replacements in the lepton spinors.}

\begin{figure}[t]
    \centering
    \includegraphics[width=0.7\linewidth]{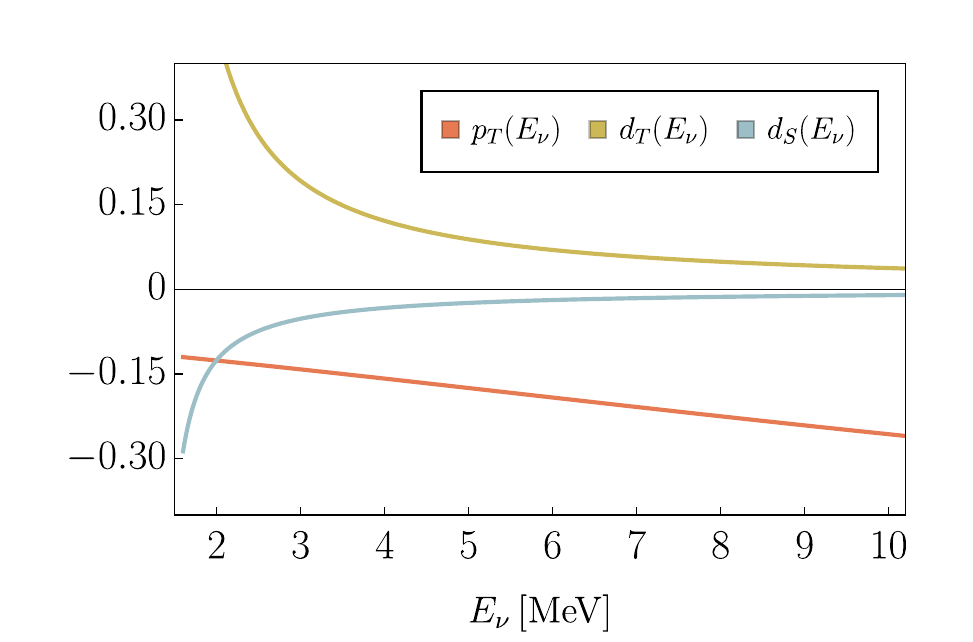}
    \caption{Dependence of $p_T(E_\nu)$, $d_T(E_\nu)$, and $d_S(E_\nu)$ functions on the neutrino energy $E_\nu$ over the relevant range. The corresponding analytical definitions are given in Eq.~\eqref{eq:pX_functions} and in Eq.~\eqref{eq:d_param_def}.}
    \label{fig:p_T-d_T-d_S}
\end{figure}

The squared amplitude for a single nuclear decay follows from Eq.~\eqref{eq:prod-ampl} as
\begin{equation}\label{eq:prod-ampl-sq}
{\scalebox{0.95}{$
    \begin{alignedat}{2}
        \cM^P_{\alpha \rho} \, \Bar{\cM}^P_{\alpha\eta} = 
        \delta_{\alpha\rho}\,\delta_{\eta\alpha}\,|A^P_L|^2 
        + \sum_X \left( [\epsilon_{X}]_{\alpha\rho}\,\delta_{\eta\alpha}\, A^P_X \, \Bar{A}^P_L + {\rm h.c.}
        \right)
        + \sum_{X,Y} [\epsilon_{X}]_{\alpha\rho}\, [\epsilon_{Y}^\dagger]_{\eta\alpha} \,A^P_X \,\Bar{A}^P_Y\,.
    \end{alignedat}
$}}
\end{equation}
Substituting the above expression into the flux formula, Eq.~\eqref{eq:mass-to-flavor}, we obtain the generalized flux in the flavor basis in matrix form:
\begin{eqnarray}\label{eq:generalized-flux}
    \frac{d\Phi^\alpha}{dE_\nu} &=& \frac{d\phi^\alpha_{\sscript{SM}}}{dE_\nu} \,\sum_{X,Y}\,\left[ \Pi^\alpha + p_{XL}\,  (\Pi^\alpha \epsilon_X) + p_{XL}^* \,(\epsilon_X^\dagger\Pi^\alpha) + p_{XY}\, (\epsilon_Y^\dagger \Pi^\alpha \epsilon_X)  \right]\,. 
\end{eqnarray}
In the above, we identified the SM (differential) flux and the production coefficients as 
\begin{eqnarray}\label{eq:SM-flux}
    \frac{d\phi^\alpha_{\sscript{SM}}}{dE_\nu} = \frac{N_S}{8\pi L^2 m_S} \int d \Pi_{P^{\prime}} |A^P_L|^2\,,
    \qquad
    p_{XY} = \cfrac{\int d \Pi_{P^{\prime}} A^P_X \, \Bar{A}^P_Y}{\int d \Pi_{P^{\prime}} |A^P_L|^2}\,,
\end{eqnarray}
where $m_S$ and $N_S$ denote the mass and number of particles of the parent nucleus $\cN$, respectively. Since antineutrinos are produced through a wide variety of beta-decay processes in the reactor, the above expressions implicitly include a sum over contributions from different $\cN - \cN^\prime$ transitions. At leading order in the LEFT coefficients, only the $p_{XL}$ coefficients contribute. Hence, in the remainder of the discussion, we omit the subscript $L$ and write $p_X$ instead of $p_{XL}$. Moreover, while Eq.~\eqref{eq:A-X-P} includes both Fermi and Gamow--Teller contributions, the latter dominate the reactor antineutrino spectrum~\cite{Hayes:2016qnu}. Therefore, in a simplified reactor setup, we neglect the contributions from Fermi transitions, and the coefficients become
\begin{eqnarray}\label{eq:pX_functions}
    p_{L} = 1\,,\qquad
    p_{R} = -1\,,\qquad
    p_{S} \approx 0 \,,\qquad
    p_{P} \approx 0 \,,\qquad
    p_{T} = -0.094-0.016\,E_\nu\,.
\end{eqnarray}
The $p_T$ coefficient acquires a non-trivial energy dependence through the underlying beta-decay dynamics in the reactor (see Fig.~\ref{fig:p_T-d_T-d_S}). Following the treatment of Ref.~\cite{Falkowski:2019xoe}, this dependence can be computed by averaging over the relevant beta-decay processes. However, in the energy range $E_\nu \in [1.5,\,10]\,\mev$ relevant for reactor neutrinos, the resulting function is well approximated by a linear form.

%%%%%%%%%%%%%%%%%%%%%%%%%%%%%%%%%%%%%%%%%%%%%%%%%%%%%%%%%%%%%%%%
%%%%%%%%%%%%%%%%%%%%%%%%%%%%%%%%%%%%%%%%%%%%%%%%%%%%%%%%%%%%%%%%

%%%%%%%%%%%%%%%%%%%%%%%%%%%%%%%%%%%%%%%%%%%%%%%%%%%%%%%%%%%%%%%%
\subsection{Detection: Inverse Beta Decay}\label{sec:detection}
%%%%%%%%%%%%%%%%%%%%%%%%%%%%%%%%%%%%%%%%%%%%%%%%%%%%%%%%%%%%%%%%
Reactor antineutrinos are typically identified via inverse beta decay (IBD) on the free protons of the liquid scintillator, as the interaction threshold for carbon nuclei significantly exceeds the energy range of the reactor flux. The detection amplitude $\cM^D_{\beta \rho}$ $\equiv$ $\cM (\bar{\nu}_\rho\, p^+ \rightarrow \ell^+_\beta n)$ can be written as
\begin{equation}\label{eq:det-ampl}
    \cM^D_{\beta \rho} 
    = \delta_{\rho\beta}\,A^D_L + \sum_{X}\, [\epsilon_{X}^\dagger]_{\rho\beta}\, A^D_X\,.
\end{equation}
The amplitudes $A_X^D$ can be obtained from the general expressions in Eqs.~\eqref{eq:A-X-P}--\eqref{eq:M-F} by noting that the IBD process involves a proton in the initial state and a neutron in the final state, implying $M_F = 1$, $r = \sqrt{3}$, $J = 1/2$ and ${\cal T}^k_{1/2} = \frac{1}{2}\sigma^k$. Finally, using the appropriate Dirac spinors for the leptons, we obtain
\begin{eqnarray}\label{eq:A-X-D}
    A^D_{L/R} &=& -\cfrac{2\,\sqrt{E_p\,E_n}\, V_{ud}}{v^2} \left[\,\delta_{J_z^\prime J_z} (\bar{v}_{\nu_\rho}\gamma^0 P_L v_{\ell_\beta} ) \,\mp\, g_A\,[\sigma^k]^{J_z}_{J_z^\prime}\, (\bar{v}_{\nu_\rho}\gamma^k P_L v_{\ell_\beta} ) \right] \,, 
    \nonumber
    \\[3pt] 
    A^D_S &=& -\cfrac{2\,\sqrt{E_p\,E_n}\, V_{ud}}{v^2}\,g_S\,\delta_{J_z^\prime J_z} ( \bar{v}_{\nu_\rho}P_R v_{\ell_\beta} )\,,
    \qquad\qquad A^D_P = 0 \,,
    \nonumber
    \\[3pt]
    A^D_T &=& -\cfrac{2\,\sqrt{E_p\,E_n}\, V_{ud}}{v^2}\, g_T \,[\sigma^k]^{J_z}_{J_z^\prime}\, (\bar{v}_{\nu_\rho}\gamma^0\gamma^k P_R v_{\ell_\beta} )\,,
\end{eqnarray}
The squared amplitude can then be written as
\begin{equation}\label{eq:det-ampl-sq}
{\scalebox{0.94}{
$\begin{alignedat}{2}
\cM^D_{\beta \rho} \, \Bar{\cM}^D_{\beta\eta} = \delta_{\rho\beta}\,\delta_{\beta\eta}\,|A^D_L|^2 
+ \sum_X \left( [\epsilon_{X}^\dagger]_{\rho\beta}\,\delta_{\beta\eta}\, A^D_X \, \Bar{A}^D_L + {\rm h.c.} 
\right)
+ \sum_{X,Y} [\epsilon_{X}^\dagger]_{\rho\beta}\, [\epsilon_{Y}]_{\beta\eta} \,A^D_X \,\Bar{A}^D_Y\,.
\end{alignedat}$
}}
\end{equation}
Substituting the above expression into the cross-section formula, Eq.~\eqref{eq:mass-to-flavor2}, we obtain
\begin{eqnarray}\label{eq:eff-cross-sec}
    \Sigma^\beta = \sigma^\beta_{\rm SM} \,\sum_{X, Y} \left[ \Pi^\beta + d_{XL}\,( \epsilon_X^\dagger \, \Pi^\beta ) + d^*_{XL}\,( \Pi^\beta \,\epsilon_X ) + d_{XY}\, ( \epsilon_X^\dagger \, \Pi^\beta \, \epsilon_Y) \right]~,
\end{eqnarray}
in the flavor basis. The SM cross-section and detection coefficients are identified as 
\begin{eqnarray}\label{eq:SM-cross-section}
    \sigma^\beta_{\sscript{SM}} = \frac{1}{4 m_p E_\nu} \int d \Pi_{D}\, |A^D_L|^2, 
    \qquad 
    d_{XY} = \frac{\int \Pi_{D} A^D_X \, \Bar{A}^D_Y}{\int \Pi_{D} |A^D_L|^2}\,.
\end{eqnarray}
As in the production case, only $d_{XL}$ coefficients are relevant up to leading order in the LEFT expansion. Hence, we drop the subscript $L$ and simply write $d_X$ instead of $d_{XL}$. These coefficients can then be evaluated as \cite{Falkowski:2019xoe}
\begin{equation}\label{eq:d_param_def}
{\scalebox{0.98}{
$\begin{alignedat}{2}
d_{L} = 1\,,\quad
d_{R} = -\frac{3g_A^2 - 1}{3g_A^2 + 1}\,,\quad
d_{S} = -\frac{g_S}{3g_A^2 + 1}\frac{m_e}{E_\nu - \Delta}\,,\quad
d_{T} = \frac{3 g_A\, g_T}{3g_A^2 + 1}\frac{m_e}{E_\nu - \Delta}\,,
\end{alignedat}$
}}
\end{equation}
whereas the pseudo-scalar coefficient $d_P$ vanishes. Here $\Delta \equiv m_n - m_p \approx 1.29\,\mev$ is the neutron-proton mass splitting and $m_e \approx 0.511\,\mev$ is the positron mass. Using the numerical values of the nucleon charges given in Sec.~\ref{sec:LY_lagr_EFT}, one finds that $d_R\approx-0.66$, while the scalar and tensor coefficients depend explicitly on the neutrino energy, with their functional dependence displayed in Fig.~\ref{fig:p_T-d_T-d_S}.

%%%%%%%%%%%%%%%%%%%%%%%%%%%%%%%%%%%%%%%%%%%%%%%%%%%%%%%%%%%%
\subsection{Survival Probability}
\label{sec:surv_prob}
%%%%%%%%%%%%%%%%%%%%%%%%%%%%%%%%%%%%%%%%%%%%%%%%%%%%%%%%%%%%
We now turn our attention to the transition probability, $\widetilde P_{\alpha\beta}$, defined by Eq.~\eqref{eq:P_eff}, which we restate here for convenience:\footnote{We refer here to the {\emph{effective}} probability $\widetilde P_{\alpha\beta}$, rather than the actual probability $P_{\alpha\beta}$ (see Sec.~\ref{sec:probability}). For brevity, the qualifier ``effective" is hereafter omitted.}
\begin{eqnarray}\label{eq:rate-and-prob}\nonumber
    R_{\alpha\beta} = \cfrac{d\phi^\alpha_\sscript{SM}}{dE_\nu}\, \widetilde P_{\alpha\beta}(L, E_\nu)\,\sigma^\beta_\sscript{SM}\,.
\end{eqnarray}
Employing Eq.~\eqref{eq:compact-rate} for the rate, we obtain
\begin{eqnarray}\label{eq:compact-rate-and-prob_alt}
    \widetilde P_{\alpha\beta}(L, E_\nu) = \Tr\left[ F\,\cfrac{d\Phi^\alpha}{dE_\nu}\,F^\dagger\,\Sigma^\beta\right]\,\left(\cfrac{d\phi^\alpha_\sscript{SM}}{dE_\nu}\right)^{\eminus1}\,(\sigma^\beta_\sscript{SM})^{\eminus1}\,.
\end{eqnarray}
Expanding to linear order in $\epsilon_X$, we obtain
\begin{align}\label{eq:trans-prob-traces}
\widetilde P_{\alpha\beta}(L, E_\nu)
&=\Tr\left[F\,\Pi^\alpha\,F^\dagger\,\Pi^\beta \right]
+\sum_X\Tr\left[ p_{X} F\,\Pi^\alpha\,\epsilon_X\,F^\dagger\,\Pi^\beta
+d_{X} \,F\,\Pi^\alpha\,F^\dagger\,\epsilon_X^\dagger\,\Pi^\beta
+ {\rm c.c.}\right]\nonumber
\\[3pt]
&=|F_{\beta\alpha}|^2
+\sum_X\,\left( p_{X}\,F_{\beta\alpha}\,[\epsilon_X F^\dagger]_{\alpha\beta}+d_{X}\,F_{\beta\alpha}\,[F^\dagger \epsilon_X^\dagger]_{\alpha\beta} \,+\, {\rm c.c.}\right)\,,
\end{align}
where we used the generalized flux and cross section in Eqs.~\eqref{eq:generalized-flux} and \eqref{eq:eff-cross-sec}. As discussed in Sec.~\ref{sec:vacuum}, in the limit of negligible matter effects, the evolution operator in the flavor basis is given by
\begin{equation}\label{eq:F-in-terms-of-U}
F_{\beta\alpha} = U_{\beta \ell}\,\exp\left(i\frac{m_\ell^2 L}{2 E_\nu}\right)\delta_{\ell m}\,U^*_{\alpha m}\,.
\end{equation}
Inserting this expression into Eq.~\eqref{eq:compact-rate-and-prob_alt}, the transition probability can be written as~\cite{Falkowski:2019kfn,Giunti:2007ry}
\begin{equation}\label{eq:trans-prob-compact}
    \widetilde P_{\alpha\beta}(L, E_\nu) = \sum_{k,\ell}\, \exp\left(-i\cfrac{\Delta m_{k\ell}^2L}{2E_\nu}\right)\, C^{\alpha\beta}_{k\ell}\,,
\end{equation}
where $C^{\alpha\beta}_{k\ell}$ are defined as
\begin{equation}
    C^{\alpha\beta}_{k\ell}=\cfrac{d\Phi^\alpha_{\ell k}}{dE_\nu}~\Sigma^\beta_{k\ell}\left(\cfrac{d\phi^\alpha_\sscript{SM}}{dE_\nu}\,\sigma^\beta_\sscript{SM}\right)^{\eminus1}\,.
\end{equation}
Expanding the coefficients $C^{\alpha\beta}_{k\ell}$ to linear order in $\epsilon_X$, one finds
\begin{equation}
\begin{alignedat}{2}
C^{\alpha\beta}_{k\ell}
=U_{\alpha k}\,U_{\beta k}^*\,U_{\alpha \ell}^*\,U_{\beta \ell}
&+p_{X} [\epsilon_X U]_{\alpha k}\,U_{\beta k}^*\,U_{\alpha \ell}^*\,U_{\beta \ell}+p^*_{X} U_{\alpha k}\,U^*_{\beta k}\,[\epsilon_X U]^*_{\alpha \ell}\,U_{\beta \ell}
\\[3pt]
&+d_{X} U_{\alpha k}\,[\epsilon_X U]^*_{\beta k}\,U^*_{\alpha \ell}\,U_{\beta \ell}
+d^*_{X} U_{\alpha k}\,U^*_{\beta k}\,U^*_{\alpha \ell}\,[\epsilon_X U]_{\beta \ell}\,,
\end{alignedat}
\end{equation}
where an implicit sum over $X=L,R,S,T$ is understood. 

In this work, we focus on the survival probability which, using the hermiticity property $C_{k\ell}^{\alpha\alpha} = (C_{\ell k}^{\alpha\alpha})^*$, can be written as
\begin{align}\label{eq:survivalPwithCs}
        \widetilde P_{\alpha\alpha}(L,E_\nu)& =C^{\alpha\alpha}_{11}+C^{\alpha\alpha}_{22}+C^{\alpha\alpha}_{33}+2\re(C_{21}^{\alpha\alpha}+C_{31}^{\alpha\alpha}+C_{32}^{\alpha\alpha})
        \nonumber\\[3pt]
        &-4\left[
        \re(C_{21}^{\alpha\alpha})\sin^2\lzm\frac{\varphi_{21}}{2}\dzm+\re(C_{31}^{\alpha\alpha})\sin^2\lzm\frac{\varphi_{31}}{2}\dzm
        +\re(C_{32}^{\alpha\alpha})\sin^2\lzm\frac{\varphi_{32}}{2}\dzm\right]
        \nonumber\\[3pt] &+2\Big[\im(C_{21}^{\alpha\alpha})\sin(\varphi_{21})+\im(C_{31}^{\alpha\alpha})\sin(\varphi_{31})+\im(C_{32}^{\alpha\alpha})\sin(\varphi_{32})\Big]\,,
\end{align}
where we introduce the shorthand $\varphi_{ij}=\Delta m_{ij}^2L/2E_\nu$. 

%%%%%%%%%%%%%%%%%%%%%%%%%%%%%%%%%%%%%%%%%%%%%%%%%%%%%%
\subsubsection*{Medium-Baseline Reactor Experiments}
%%%%%%%%%%%%%%%%%%%%%%%%%%%%%%%%%%%%%%%%%%%%%%%%%%%%%%
For reactor experiments, the measured event rates are governed by the electron antineutrino survival probability, corresponding to $\alpha=e$. In medium-baseline facilities, with $L\sim 50\,\rm{km}$, the oscillation pattern is primarily sensitive to the solar parameters $\Delta m_{21}^2$ and $\theta_{12}$. As a consequence, NSI contributions suppressed by powers of the reactor angle $s_{13}$ can be safely neglected, since $s_{13}$ is numerically small. The effective probability then takes the form
\begin{equation}\label{eq:sur_prob_e_JUNO_w_NSI}
{\scalebox{0.98}{
$
\begin{alignedat}{2}
\widetilde P_{ee}(L,E_\nu)= ~\kappa_{e}
\times
\bigg[1 
&-\lzm 4s_{12}^2c_{12}^2c_{13}^4 +\,F(E_\nu)\dzm \sin^2\lzm\frac{\Delta m_{21}^2 L}{4E_\nu}\dzm 
+\,G(E_\nu)\sin\lzm\frac{\Delta m_{21}^2 L}{2E_\nu}\dzm
\\[4pt]
&-4c_{12}^2s_{13}^2c_{13}^2\sin^2\lzm\frac{\Delta m_{31}^2 L}{4E_\nu}\dzm -4s_{12}^2s_{13}^2c_{13}^2\sin^2\lzm\frac{\Delta m_{32}^2L}{4E_\nu}\dzm
\bigg] + \cO(\epsilon_X^2)\,,
\end{alignedat}
$
}}
\end{equation}
where 
\begin{eqnarray}\label{eq:define-F-G}
F(E_\nu) &\equiv& 4\re \left( C^{ee}_{21} - [C^{ee}_{21}]_{\sscript{SM}}\right) = \,4s_{12}c_{12}c_{13}^3 \,(c_{12}^2-s_{12}^2)\,\sum_X(p_X+d_X)\,\re[\widetilde X]\,, 
\nonumber
\\[3pt]
G(E_\nu) &\equiv& 2\, \im(C^{ee}_{21}) = 2s_{12}c_{12}c_{13}^3\sum_X(p_X-d_X)\,\im[\widetilde X]
\end{eqnarray}
encapsulate the leading-order contributions from NSIs. We define the linear combination of NSI parameters
\begin{align}\label{eq:def_tilde_X}
[\widetilde X]=c_{23}[\epsilon_X]_{e\mu}-s_{23}[\epsilon_X]_{e\tau}\,,    
\end{align}
with $s_{23}\approx0.68$ and $c_{23}\approx0.73$. The factor $\kappa_e\equiv 1+2(p_X + d_X)\re[\epsilon_X]_{ee}$, appearing in the first line of Eq.~\eqref{eq:sur_prob_e_JUNO_w_NSI}, contains the flavor-diagonal NSI contributions and only modifies the overall normalization of the survival probability, without introducing new oscillatory structures. Moreover, $\kappa_e$ cancels in the ratio of near-to-far event rates at fixed $E_\nu$, rendering it unobservable in the oscillation analysis. We refer the reader to Ref.~\cite{Falkowski:2019xoe} for a detailed discussion of these terms. Since they are strongly constrained by precision measurements of nuclear and hadronic decays in which the neutrino is not detected~\cite{Falkowski:2020pma}, we do not consider them further. 

Several additional remarks regarding Eq.~\eqref{eq:sur_prob_e_JUNO_w_NSI} are in order:
\begin{itemize}
    \item For a medium-baseline reactor neutrino experiment, the survival probability is primarily driven by the oscillation term in the first line of Eq.~\eqref{eq:sur_prob_e_JUNO_w_NSI}, which is governed by the solar mass-squared splitting $\Delta m^2_{21}$. The second line gives rapid, small-amplitude oscillations superimposed on the main solar oscillation wave. An excellent energy resolution would allow one to resolve the phase difference that appears in these fast oscillations depending on the mass ordering (normal vs inverted)~\cite{JUNO:2022mxj,JUNO:2024jaw}. Otherwise, one only measures an averaged, non-zero effect that is identical for both orderings. Furthermore, NSI effects associated with these high-frequency modes have been neglected here, as they would constitute mere corrections to an already subleading term.
    \item We note that Eq.~\eqref{eq:sur_prob_e_JUNO_w_NSI} is not valid for short-baseline reactor experiments (with $L \sim O(1)$ km) such as Daya Bay. These experiments probe oscillations driven by $\Delta m_{31}^2$, for which the leading contribution to the survival probability is governed by $\theta_{13}$. Hence, the leading nonstandard contributions are those proportional to $s^n_{13}$, 
    which are neglected in Eq.~\eqref{eq:sur_prob_e_JUNO_w_NSI}. For the study of those experiments one should employ different approximations, see Ref.~\cite{Falkowski:2019xoe} for further details. For a combined JUNO–Daya Bay analysis one would simply use the general expression in Eq.~\eqref{eq:survivalPwithCs}, which is valid for both experiments.
    \item The $[\widetilde X]$ structures defined in Eq.~\eqref{eq:def_tilde_X} do not depend on the CP-violating phase $\delta_\sscript{CP}$. In principle, $\delta_\sscript{CP}$ can enter the survival probability through other combinations of NSI parameters. However, such contributions appear only in terms suppressed by $s^n_{13}$ 
    and are therefore subleading. This differs from short-baseline reactor experiments such as Daya Bay, which probe a different linear combination of NSI parameters, given by $[X] = e^{i\delta_\sscript{CP}}\lzm s_{23}[\epsilon_X]_{e\mu} + c_{23}[\epsilon_X]_{e\tau}\dzm$, in which $\delta_\sscript{CP}$ enters already at leading order~\cite{Falkowski:2019xoe}.
    \item Matter effects become non-negligible for long baselines and high-precision medium baselines. Assuming a constant matter-density profile, which is often a sufficiently accurate approximation, the survival probability in Eq.~\eqref{eq:sur_prob_e_JUNO_w_NSI} retains the same form as in vacuum, with the mass splittings and mixing angles replaced by their effective parameters in matter, which depend on the density and NC (non-)standard interactions. For the non-constant density profile, this evolution must be treated numerically in the usual way (see Sec.~\ref{sec:matter}).
    \item We note that our result in Eq.~\eqref{eq:sur_prob_e_JUNO_w_NSI} for medium baselines and the short-baseline result of Ref.~\cite{Falkowski:2019xoe} do not exactly match those of Ref.~\cite{Chaves:2021kxe}, even after taking into account the different PMNS parametrization used in that work (and the consequent differences in the definitions of the Wilson coefficients).
\end{itemize}
Based on Eq.~\eqref{eq:sur_prob_e_JUNO_w_NSI}, we can identify the combinations of NSI parameters that can be constrained. 
Let us start by pointing out that the first term can be rewritten as
\begin{eqnarray}
4 \,\re(C^{ee}_{21})
&=& 4s_{12}^2c_{12}^2c_{13}^4 + F(E_\nu)\nonumber\\[4pt]
&=& \cos^4\theta_{13}\Big[\sin^2 2\theta_{12} + \frac{\sin 4\theta_{12}}{\cos\theta_{13}} \sum_X (p_X+d_X)\re[\widetilde X]\Big]
\nonumber\\[3pt]
&\approx&\cos^4\theta_{13}\sin^2 \Big[2\theta_{12} + \sum_X (p_X+d_X)\re[\widetilde X] \Big]\,,
\end{eqnarray}
where in the last step we assume that $(p_X + d_X) \re[\widetilde X] \ll 2\theta_{12}$, and omit NSI terms suppressed by $s_{13}$. As a result, any energy-independent contribution to $F(E_\nu)$ affects the survival probability only through a shift of the mixing angle $\theta_{12}$. This is precisely the case for $[\widetilde L]$ and $[\widetilde R]$. Consequently, these effects cannot be disentangled from $\theta_{12}$ using reactor oscillation data alone and would require an independent determination of $\theta_{12}$ from other experiments.\footnote{This argument is not exact, as it neglects NSI contributions to the rapid oscillations, which are suppressed by powers of $\theta_{13}$ and also depend on $\theta_{12}$.} A similar result was found in the $\theta_{13}$ extraction from short-baseline reactor data~\cite{Falkowski:2019xoe}. On the other hand, the contribution proportional to $\im[\widetilde L]$ vanishes, since $p_L=d_L=1$ and therefore $G(E_\nu)=0$. 
Finally, as discussed in Sec.~\ref{sec:EFT_ladder}, pseudoscalar interactions do not appear at leading order in the recoil expansion.

The above considerations imply that the independent linear combinations of NSI parameters that can be constrained are $\im[\widetilde R]$, $\re[\widetilde S]$, $\im[\widetilde S]$, $\re[\widetilde T]$ and $\im[\widetilde T]$. Their relative sensitivities can already be anticipated from the interplay between the production and detection coefficients and the trigonometric structures multiplying the corresponding NSI contributions. In particular, the sizable $(p_R-d_R)$ and $(p_T-d_T)$ combinations, together with the comparatively less suppressed oscillatory structure accompanying the imaginary parts, suggest that $\im[\widetilde R]$ and $\im[\widetilde T]$ should yield the strongest constraints.

%%%%%%%%%%%%%%%%%%%%%%%%%%%%%%%%%%%%%%%%%%%%%%%%%%%%%%%%%%%%%%%%%%%%%%%%%%%%%%%%%%%%
\section{Phenomenology at JUNO}
\label{sec:pheno_at_JUNO}
%%%%%%%%%%%%%%%%%%%%%%%%%%%%%%%%%%%%%%%%%%%%%%%%%%%%%%%%%%%%%%%%%%%%%%%%%%%%%%%%%%%%

Having developed the general framework for reactor neutrino experiments in Sec.~\ref{sec:JUNO-th}, we now specialize to the medium-baseline JUNO experiment. JUNO (Jiangmen Underground Neutrino Observatory) is a multipurpose medium-baseline reactor neutrino experiment located in Kaiping, China. The currently available dataset, corresponding to 59.1 days of operation, already enables high-precision measurements of the solar mixing angle $\theta_{12}$ and the mass-squared difference $\Delta m_{21}^2$~\cite{JUNO:2025gmd}.

As an implementation of the detection principles outlined in Sec.~\ref{sec:detection}, JUNO utilizes a massive liquid scintillator detector to identify these antineutrinos via IBD. The experiment receives reactor antineutrinos primarily from the Yangjiang and Taishan nuclear power plants, which host a total of eight reactor cores (six at Yangjiang and two at Taishan) at an average baseline of approximately 52.5 km from the detector. In addition, a smaller contribution arises from more distant reactors, with the dominant component originating from the Daya Bay complex. In our analysis, these contributions are taken into account through their relative flux weights, which encode the fractional contribution of each reactor core to the total flux. The corresponding reactor configurations and the flux weights are summarized in Tab.~\ref{tab:JUNO_base_flux}.

%%%%%%%%%%%%%%%%%%%%%%%%%%%%%%%%%%%%%%%%%
%%%%%%%%%%%%%%%%%%%%%%%%%%%%%%%%%%%%%%%%%
\begin{table*}[t]
\centering
\scalebox{0.95}{
\begin{tabular}{l@{\hspace{2.5cm}}c@{\hspace{2.5cm}}c}
\toprule
\textbf{Reactor}
& \textbf{Baseline $\bm{(\text{km})}$}
& \textbf{Flux weight $\bm{(\%)}$}
\\
\midrule
%%%%%%%%%%%%%%%%%%%%%
\textbf{Taishan} & $52.71$ & $32.1$ \\
\quad Core 1     & $52.77$ & $16.0$ \\
\quad Core 2     & $52.64$ & $16.1$ \\
\addlinespace[4pt]
\cdashline{1-3}[.4pt/2pt]
\addlinespace[4pt]
\textbf{Yangjiang} & $52.46$ & $61.5$ \\
\quad Core 1       & $52.74$ & $10.1$ \\
\quad Core 2       & $52.82$ & $10.1$ \\
\quad Core 3       & $52.41$ & $10.3$ \\
\quad Core 4       & $52.49$ & $10.2$ \\
\quad Core 5       & $52.11$ & $10.4$ \\
\quad Core 6       & $52.19$ & $10.4$ \\
\addlinespace[4pt]
\cdashline{1-3}[.4pt/2pt]
\addlinespace[4pt]
\textbf{Daya Bay} & $215$ & $6.4$ \\
%%%%%%%%%%%%%%%%%%%%%
\addlinespace[4pt]
\bottomrule
\end{tabular}
}
\caption{Summary of the reactor cores contributing to the JUNO antineutrino flux, including their baselines $L_r$ and relative flux weights $w_r$, taken from Ref.~\cite{JUNO:2022mxj}.}
\label{tab:JUNO_base_flux}
\end{table*}
%%%%%%%%%%%%%%%%%%%%%%%%%%%%%%%%%%%%%%%%%
%%%%%%%%%%%%%%%%%%%%%%%%%%%%%%%%%%%%%%%%%

%%%%%%%%%%%%%%%%%%%%%%%%%%%%%%%%%%%%%%%%%%%%%%%%%%%%%%%%%%%%%%%%%%%%%%%%%%%%%%%%%%%%
\subsection{Construction of $\chi^2$}
%%%%%%%%%%%%%%%%%%%%%%%%%%%%%%%%%%%%%%%%%%%%%%%%%%%%%%%%%%%%%%%%%%%%%%%%%%%%%%%%%%%%
The JUNO measurement relies on the reconstructed prompt-energy spectrum of the reactor antineutrino events, with the prompt energy defined as the visible energy deposited by the positron in inverse beta decay~\cite{JUNO:2025gmd,Esteban:2026phq}. The measured spectrum is given as the number of observed events in bins of the reconstructed prompt energy $E^\sscript{pr}_i$, corresponding to the total event rate $N_i^{\sscript{S+B}}$, which includes both signal ($N_i^\sscript{S}$) and background ($N_i^\sscript{B}$) contributions. 

For the construction of the $\chi^2$ function we use the information presented in the JUNO study~\cite{JUNO:2025gmd}. In the absence of publicly available JUNO data, the measured spectrum, the corresponding unoscillated signal spectrum, and the central values of the background contributions are obtained by digitizing Fig.~3 of Ref.~\cite{JUNO:2025gmd} (see App.~\ref{app:data} for the numerical values obtained). This strategy allows us to directly extract the relevant experimental inputs without explicitly implementing a detailed modeling of the reactor flux~\cite{JUNO:2025gmd,Esteban:2026phq}. As we demonstrate below, this simple approach provides a numerically stable and accurate implementation of the experimental setup, successfully reproducing the official JUNO results, and thereby justifying its use in the NSI analysis.

For each of the 65 digitized bins, labeled by $i$, the signal contribution is defined as
\begin{equation}
    N_i^\sscript{S}(\vec\xi\,)=N_i^{\sscript{S+B}}-N_i^\sscript{B}(\vec\xi\,)\,,
\end{equation}
where $\vec\xi$ collectively denotes the nuisance parameters affecting the background and, hence, the signal. The background contribution $N_i^\sscript{B}(\vec\xi\,)$ consists of the five components considered in the JUNO analysis~\cite{JUNO:2025gmd}: geoneutrinos ($N_i^\sscript{Geo}$), $^9\mathrm{Li}/^8\mathrm{He}$ from cosmic muon spallation ($N_i^\sscript{LiHe}$), $^{214}\mathrm{Bi}/^{214}\mathrm{Po}$ from radon decay ($N_i^\sscript{BiPo}$), the reactor antineutrinos from distant reactors ($N_i^\sscript{world}$), and a residual contribution from other sources ($N_i^\sscript{other}$). The corresponding systematic uncertainties are incorporated through six nuisance parameters, implemented via Gaussian pull terms:
\begin{equation}\label{eq:N_bckg}
    \begin{alignedat}{2}
    N_i^\sscript{B}(\vec\xi\,)&=\sum_a (1+\xi_a)\,N_i^a
    +\lzm 1+\xi_\sscript{LiHe}^{(1)}+\xi_\sscript{LiHe}^{(2)}\,\frac{E_i^\sscript{pr}}{\mev} \dzm\,N_i^\sscript{LiHe}\,,
    \end{alignedat}
\end{equation}
where $a=\{\rm Geo, LiHe, BiPo, World, Other\}$ labels the background components. To each nuisance parameter $\xi_a$ we assign Gaussian widths corresponding to normalization uncertainties $\sigma_{\xi_a}$ of $42\%$, $33\%$, $56\%$, $10\%$, and $100\%$ for $\xi_\sscript{Geo}$, $\xi_\sscript{LiHe}^{(1)}$, $\xi_\sscript{BiPo}$, $\xi_\sscript{world}$, and $\xi_\sscript{other}$, respectively \cite{JUNO:2025gmd,Esteban:2026phq}. In addition, the shape uncertainty of the $^9\mathrm{Li}/^8\mathrm{He}$ background is modeled by an extra nuisance parameter $\xi_\sscript{LiHe}^{(2)}$, taken to have a $20\%$ uncertainty at $1\,\mev$ and scaling linearly with energy. In the present analysis, we restrict ourselves to this set of nuisance parameters, which correspond to the dominant sources of systematic uncertainty and carry the largest assigned uncertainties. Additional systematics, such as energy-scale, resolution, and reactor-flux uncertainties, can be incorporated in a more complete treatment~\cite{JUNO:2025gmd,Esteban:2026phq}.

We now turn to the theoretical prediction for the signal. For the predicted number of events in bin $i$ we use
\begin{align}\label{eq:pred_signal_NP}
    N_i^\sscript{P}=C\, \sum_{r=1}^9 \,w_r\,\phi_e^\sscript{SM}(E_{\nu}^i)\,\sigma_e^\sscript{SM}(E_{\nu}^i) \, \widetilde P_{ee}(L_r,E_{\nu}^i)\,,
\end{align}
where $C$ is an overall normalization factor accounting for the number of target protons, the exposure time, and the selection efficiencies; $L_r$ and $w_r$ denote the baseline and relative flux weights, respectively (see Tab.~\ref{tab:JUNO_base_flux}), and the sum runs over all nine cores contributing to the JUNO flux. Furthermore, we approximate the neutrino energy in bin $i$ by its value at the center of the bin, $E_{\nu}^i$, which in turn is related to the prompt energy through $E_{\nu}^i = E_\sscript{pr}^i+0.78\,\mev$, where the detailed detector effects, such as energy resolution and response functions, are neglected~\cite{JUNO:2025gmd,Esteban:2026phq}.

In order to relate the predicted signal and the unoscillated spectrum reported by the JUNO collaboration, which we extract by digitizing the corresponding distribution (see App.~\ref{app:data}), we introduce the unoscillated event rate by setting $L_r=0$:
\begin{align}  
    N_i^\sscript{unosc}=C\,  \phi_e^\sscript{SM}(E_{\nu}^i)\,\sigma_e^\sscript{SM}(E_{\nu}^i) \widetilde P_{ee}(0,E_{\nu}^i)\,,
\end{align}
where we use $\sum_{r=1}^9 w_r=1$. Combining the previous two equations, the predicted signal can be expressed in terms of the unoscillated event rate as
\begin{align}\label{eq:N_pred}
    N_i^\sscript{P}=N_i^\sscript{unosc}\sum_{r=1}^9 w_r \,\frac{\widetilde P_{ee}(L_r,E_{\nu}^i)}{\widetilde P_{ee}(0\,,E_{\nu}^i)}\,,
\end{align}
where the survival probability $\widetilde P_{ee}$ is understood to depend on the neutrino oscillation parameters and, in the presence of non-standard interactions, on the NSI coefficients introduced in Sec.~\ref{sec:EFT_ladder}.

With these ingredients in place, we construct the Gaussian $\chi^2$ function over the reconstructed energy bins:\footnote{We use the Gaussian likelihood for simplicity, as it provides an accurate enough description for the purposes of this work. A more refined approach can be implemented using Poisson or CNP likelihoods~\cite{JUNO:2025gmd,Esteban:2026phq}.}
\begin{equation}\label{eq:chi2_gaussian}
    \chi^2_\sscript{JUNO}(\vec{\alpha}\,,\vec\xi\,)=\sum_{i=1}^{65}\lzm\frac{N_i^\sscript{S}(\vec\xi\,)-N_i^\sscript{P}(\vec{\alpha}\,)}{\sqrt{N_i^{\sscript{S+B}}}}\dzm^2
    \,+\,\sum_{j=1}^6\frac{\xi_j^2}{\sigma^2_{\xi_j}}
    \,+\,\chi^2(\sin^2\theta_{12},\Delta m_{21}^2)\,,
\end{equation}
where $\vec{\alpha}$ collectively denotes the PMNS parameters, $\Delta m^2_{jk}$, and NSIs. The first term in Eq.~\eqref{eq:chi2_gaussian} encodes the statistical agreement between prediction and observation in each bin, 
while the second term accounts for the Gaussian pulls associated with the background nuisance parameters. 

The last term in Eq.~\eqref{eq:chi2_gaussian} implements Gaussian pulls for the solar oscillation parameters based on the global-fit results derived in Ref.~\cite{Esteban:2024eli}, i.e., $\sin^2\theta_{12} = 0.308^{+0.012}_{-0.011}$  and  $\Delta m_{21}^2 = (7.49\pm0.19)\times\, 10^{-5}\, \textrm{eV}^2$ (see also \cite{deSalas:2020pgw}). Their inclusion accounts for the existing experimental constraints from other measurements that are sensitive to these parameters, and reduces degeneracies between standard oscillation effects and NSI contributions in the fit. This procedure assumes that the determination of these parameters in global fits is not significantly affected by new physics contributions (or at least by those relevant for JUNO). 
Assessing this assumption rigorously would require repeating the global fit in the presence of the same set of operators, which lies beyond the scope of this work and is in fact one of the ultimate goals of the EFT program to which the present analysis contributes. It is nevertheless a reasonable working assumption: charged-current NSIs enter each experiment through its specific production and detection processes, and with a dependence on the PMNS parameters and the neutrino masses that varies with the energies and baselines involved. The weakly constrained directions in parameter space are therefore expected to differ from experiment to experiment, so that they should be substantially better resolved in a global fit than in any individual measurement.
By contrast, fixing $\sin^2\theta_{13}$ and $\Delta m_{31}^2$ to their normal-ordering best-fit values has a minimal impact on the analysis compared to including them as Gaussian pull terms.

%%%%%%%%%%%%%%%%%%%%%%%%%%%%%%%%
%%%%%%%%%%%%%%%%%%%%%%%%%%%%%%%%
\begin{figure}[t]
    \centering
    \includegraphics[width=0.9\linewidth]{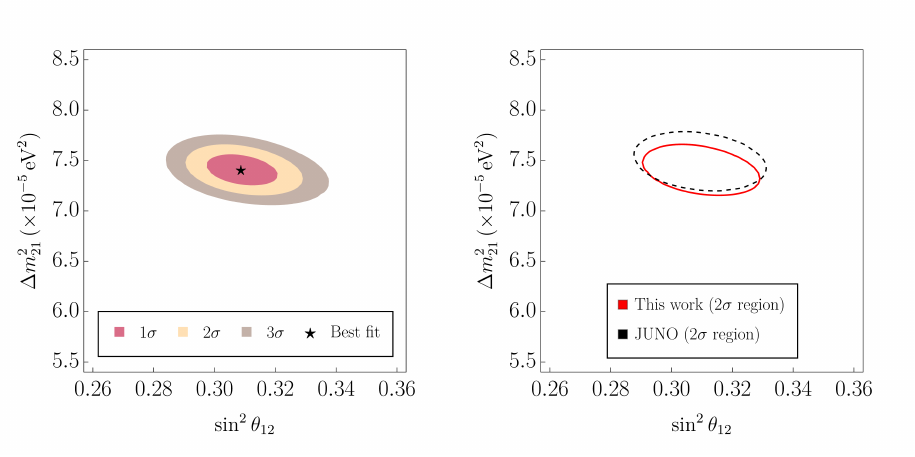}
    \caption{Validation of the analysis framework in the SM limit. \textit{Left:} $1\sigma$, $2\sigma$ and $3\sigma$ confidence regions in the $(\sin^2\theta_{12},\Delta m_{21}^2)$ plane (see Sec.~\ref{sec:validation} for details). \textit{Right:} Comparison of the $2\sigma$ confidence region obtained in this work with the corresponding contour reported by the JUNO analysis~\cite{JUNO:2025gmd}.}
    \label{fig:JUNO_validation_plot}
\end{figure}
%%%%%%%%%%%%%%%%%%%%%%%%%%%%%%%%
%%%%%%%%%%%%%%%%%%%%%%%%%%%%%%%%

%%%%%%%%%%%%%%%%%%%%%%%%%%%%%%%%%%%%%%%%%%%%%%%%%%%%%%%%%%%%%%%%%%%%%%%%%%%%%%%%%%%%
\subsection{Validation of Our Approach}\label{sec:validation}
%%%%%%%%%%%%%%%%%%%%%%%%%%%%%%%%%%%%%%%%%%%%%%%%%%%%%%%%%%%%%%%%%%%%%%%%%%%%%%%%%%%%
As a first application of the $\chi^2$ function defined in Eq.~\eqref{eq:chi2_gaussian}, we perform a validation of our framework by extracting the solar oscillation parameters $\sin^2\theta_{12}$ and $\Delta m_{21}^2$, in the absence of NSI effects. In this limit, the electron antineutrino survival probability reduces to the well-known result
\begin{equation}
    \begin{alignedat}{2}
        P^\sscript{\sscript{SM}}_{ee}(L_r,E_{\nu}^i)=1&
        -4s_{12}^2c_{12}^2c_{13}^4\sin^2\lzm\frac{\Delta m_{21}^2 L_r}{4E_{\nu}^i}\dzm
        \\[3pt]
        & -4c_{12}^2s_{13}^2c_{13}^2\sin^2\lzm\frac{\Delta m_{31}^2 L_r}{4E_{\nu}^i}\dzm
        -4s_{12}^2s_{13}^2c_{13}^2\sin^2\lzm\frac{\Delta m_{32}^2L_r}{4E_{\nu}^i}\dzm
        \,.
    \end{alignedat}
\end{equation}
In this limit, the effective and actual probabilities coincide, and we therefore omit the tilde on the left-hand side. Likewise, we omit the factor $P_{ee}(0,E)$, which reduces to unity in the absence of NSI effects.

We now substitute this expression into the $\chi^2$ given by Eq.~\eqref{eq:chi2_gaussian}. Crucially, for the purpose of this validation, the Gaussian pulls for $\sin^2\theta_{12}$ and $\Delta m_{21}^2$ are {\it not} included, ensuring that their extraction is driven entirely by the JUNO data and not by external inputs. Profiling over the background nuisance parameters then yields the $\chi^2$ as a function of $\sin^2\theta_{12}$ and $\Delta m_{21}^2$, from which the confidence regions shown in Fig.~\ref{fig:JUNO_validation_plot} are obtained. The resulting contours are in good agreement with those reported by the JUNO collaboration~\cite{JUNO:2025gmd} (see right panel in Fig.~\ref{fig:JUNO_validation_plot}), demonstrating that our simplified approach captures the relevant features of the experimental analysis.

%%%%%%%%%%%%%%%%%%%%%%%%%%%%%%%%%%%%%%%%%%%%%%%%%%%%%%%%%%%%%%%%%%%%%%%%%%%%%%%%%%%%
\subsection{Constraints on Non-Standard Interactions}
%%%%%%%%%%%%%%%%%%%%%%%%%%%%%%%%%%%%%%%%%%%%%%%%%%%%%%%%%%%%%%%%%%%%%%%%%%%%%%%%%%%%
The good agreement between our results and the JUNO analysis obtained in the previous section provides a solid basis for a consistent study of non-standard interactions. 

First, we note that the standard analysis remains valid for CP-conserving vector interactions under a straightforward reinterpretation of the mixing angle. Specifically, in this scenario, the extracted parameter is not the vacuum mixing angle $\theta_{12}$, but rather an effective angle $\widetilde\theta_{12}$ that shifts to absorb these nonstandard contributions:
\begin{align}\label{eq:shifted_theta}
\widetilde \theta_{12}= \theta_{12} + \re[\widetilde L] - \frac{3g_A^2}{3g_A^2+1}\,\re[\widetilde R]~.
\end{align}

%%%%%%%%%%%%%%%%%%%%%%%%%%%%%%%%%%%%%%%%%%%%%%%%%%%%%%%%%%%%%%%%%
%%%%%%%%%%%%%%%%%%%%%%%%%%%%%%%%%%%%%%%%%%%%%%%%%%%%%%%%%%%%%%%%%
\begin{table}[t]
\centering
\scalebox{1.0}{
\begin{tabular}{c@{\hspace{1.55cm}}c@{\hspace{1.55cm}}c@{\hspace{1.55cm}}c}
\toprule
\multirow{1}{*}{\textbf{NSI}}&\multirow{1}{*}{\textbf{Best fit}}&\multirow{1}{*}{$\bm{1\sigma}$}&\multirow{1}{*}{$\bm{2\sigma}$}
\\
\midrule
\addlinespace[0.2cm]
%%%%%%%%%%%%%%%%%%%%%%%%%%%%%%%%%%%%%%%%%%%%%%%
$\im[\widetilde R]$&$\phantom{\eminus}0.15$&$[0.07,0.24]$&$[\eminus0.02,0.33]$
\\[0.2cm]
$\re[\widetilde S]$&$\phantom{\eminus}0.41$&$[\eminus0.21,1.06]$&$[\eminus0.80,1.74]$
\\[0.2cm]
$\im[\widetilde S]$&$\phantom{\eminus}0.50$&$[\eminus0.12,1.11]$&$[\eminus0.76,1.70]$
\\[0.2cm]
$\re[\widetilde T]$&$\eminus0.49$&$[\eminus0.83,\eminus0.15]$&$[\eminus1.17,0.18]$
\\[0.2cm]
$\im[\widetilde T]$&$\phantom{\eminus}0.08$&$[\eminus0.02,0.18]$&$[\eminus0.12,0.28]$
\\[0.1cm]
\bottomrule
\end{tabular}
}

\caption{Best-fit values and corresponding $1\sigma$ and $2\sigma$ confidence intervals for the NSI combinations. In this analysis, only one NSI combination is allowed to vary at a time, while all remaining ones are fixed to zero. The standard oscillation and nuisance parameters are profiled over in the fit.}
\label{tab:One_dim_analysis_NSI}
\end{table}
%%%%%%%%%%%%%%%%%%%%%%%%%%%%%%%%%%%%%%%%%%%%%%%%%%%%%%%%%%%%%%%%%
%%%%%%%%%%%%%%%%%%%%%%%%%%%%%%%%%%%%%%%%%%%%%%%%%%%%%%%%%%%%%%%%%

\begin{figure}[t]
    \centering
    \includegraphics[width=1.00\linewidth]{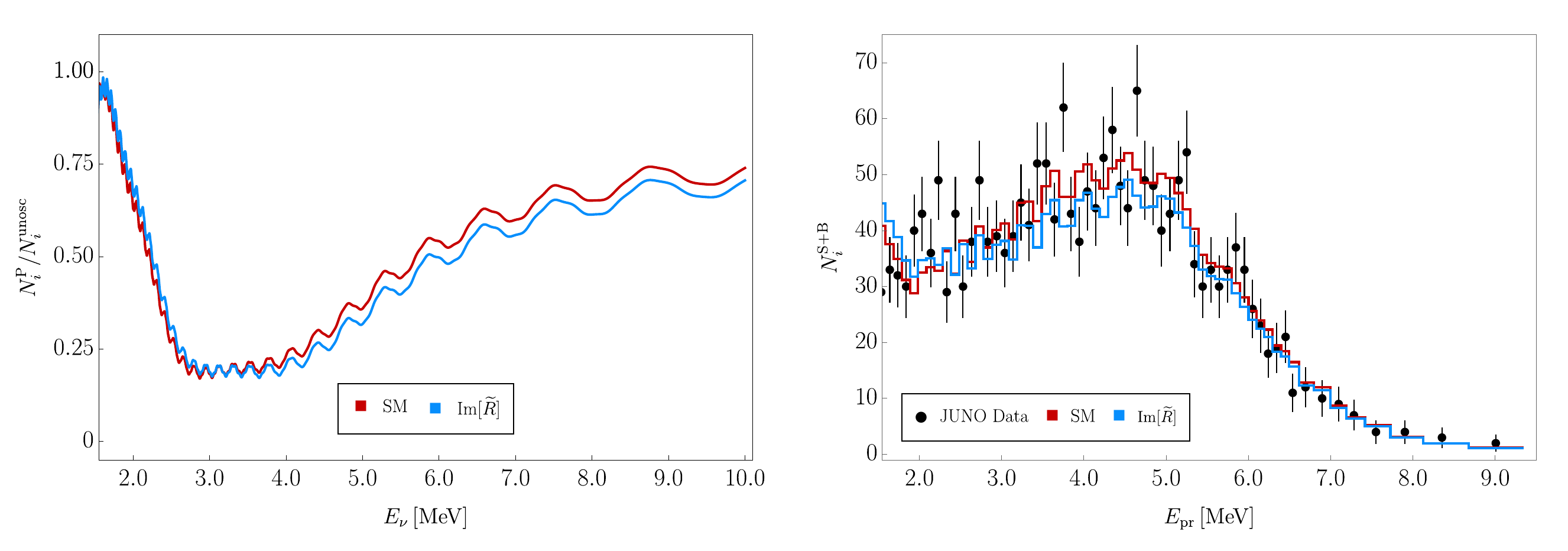}
    \caption{\textit{Left:} Far-to-near event ratio as a function of the neutrino energy $E_\nu$, shown for the SM prediction and for the NSI scenario with nonzero $\im[\widetilde R]$.  \textit{Right:} Comparison between the JUNO data $N_i^{\sscript{S+B}}$ and the predicted event rates in the SM and for the NSI scenario with nonzero $\im[\widetilde R]$. Each line in both panels is obtained using the best-fit values of the corresponding parameters (NSI, background nuisance, and standard oscillation parameters). The best-fit values differ for each line.}
\label{fig:NSI_probs}
\end{figure}

We now move to less trivial cases, where the effect cannot be absorbed into the existing parameters. For simplicity, the NSI combinations $[\widetilde X]$ are switched on individually, assuming only either the real or imaginary part is non-zero at a time. In this setup, the $\chi^2_\sscript{JUNO}$ function defined in Eq.~\eqref{eq:chi2_gaussian} depends on the chosen NSI parameter as well as on the standard oscillation parameters and the nuisance parameters describing systematic uncertainties. For each value of the NSI parameter under consideration we profile over all remaining parameters, obtaining a profiled $\chi^2(\widetilde X)$ function, from which the best-fit value and the corresponding confidence intervals can be extracted. The resulting bounds obtained from this one-parameter analysis are summarized in Tab.~\ref{tab:One_dim_analysis_NSI}. A few remarks are in order: 
\begin{itemize}
    \item  
    $\im[\tilde R]$ and $\im[\widetilde T]$ are the most tightly constrained parameters, as anticipated in the previous section. Fig.~\ref{fig:NSI_probs} illustrates the impact of $\im[\widetilde R]$ on both the weighted survival probability and the predicted event-rate spectrum by comparing the SM prediction with the corresponding best-fit NSI scenario. The remaining coefficients exhibit broader allowed regions. In any case, the bounds on the NSI parameters obtained at this stage of JUNO data taking should be regarded as indicative. They will improve with increased statistics and can be further strengthened by combining JUNO data with other experiments. 
    \item The confidence intervals for all NSI parameters are approximately symmetric around the best-fit values, with only mild deviations.
    \item One can translate the constraints on the NSI coefficients into effective scales using the matching relations of Sec.~\ref{sec:EFT_ladder}. Using the $1\sigma$ values in Tab.~\ref{tab:One_dim_analysis_NSI}, we find that the most constrained directions, such as $\im[\widetilde T]$, probe effective scales of $\Lambda\sim\mathcal{O}(700\,\gev)$, while more weakly constrained coefficients, such as $\re[\widetilde S]$ and $\im[\widetilde S]$ probe scales of $\Lambda\sim\mathcal{O}(150\,\gev)$. This highlights the range of scales probed, exposing the limitations of the SMEFT interpretation while confirming the validity of the LEFT.
    \item It is important to note that the survival probability used in our analysis is expanded only to $\cO(\epsilon_X)$, as indicated in Eq.~\eqref{eq:sur_prob_e_JUNO_w_NSI}. As a result, bounds on NSI coefficients that are not smaller than unity should be interpreted with caution. In such cases, a consistent analysis necessitates the use of the full survival probability, in order to ensure reliable results. The formalism developed in this work provides all the ingredients necessary to carry out such a non-linear analysis, including all contributions up to $\mathcal O(\epsilon_X^4)$. Nevertheless, for simplicity and in order to highlight the key phenomenological features, the present analysis is restricted to the linearized treatment.
\end{itemize}

%%%%%%%%%%%%%%%%%%%%%%%%%%%%%%%%%%%%%%%%%%%%%%%%%%%%%%%%%%%%%%%%%
%%%%%%%%%%%%%%%%%%%%%%%%%%%%%%%%%%%%%%%%%%%%%%%%%%%%%%%%%%%%%%%%%
%%%%%%%%%%%%%%%%%%%%%%%%%%%%%%%%%%%%%%%%%%%%%%%%%%%%%%%%%%%%%%%%%
%%%%%%%%%%%%%%%%%%%%%%%%%%%%%%%%%%%%%%%%%%%%%%%%%%%%%%%%%%%%%%%%%

\begin{figure}[t]
    \centering
    \includegraphics[width=1.0\linewidth]{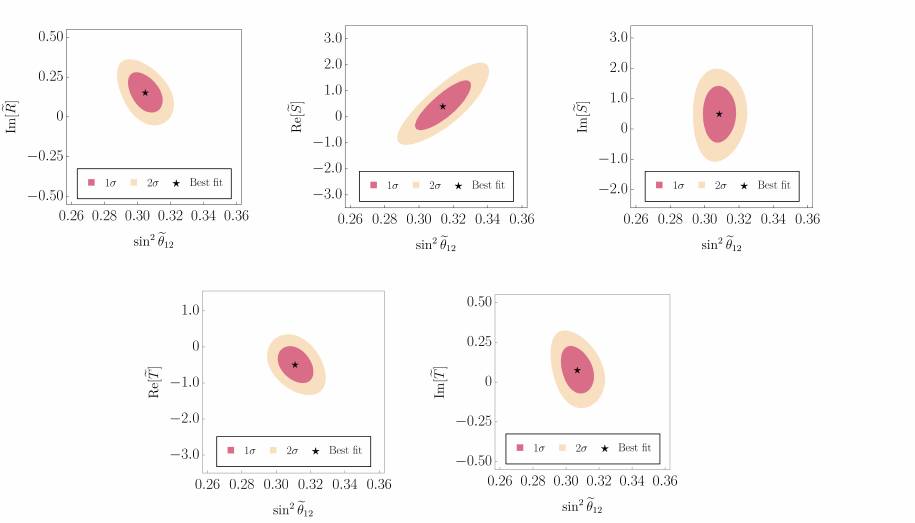}
    \caption{$1\sigma$ and $2\sigma$ confidence regions from the NSI analysis. In all panels, NSI parameters not shown are set to zero, with all remaining oscillation and nuisance parameters profiled over.}
    \label{fig:Two_dim_NSI_plot1}
\end{figure}

Next, we study the correlation between the NSI parameters and the solar mixing angle. Profiling the $\chi^2_\sscript{JUNO}$ function over all remaining oscillation and nuisance parameters yields the profile $\chi^2(\widetilde X,\sin^2\widetilde\theta_{12})$. The allowed regions in the $(\widetilde X,\sin^2\widetilde\theta_{12})$ plane are then obtained using the standard $\Delta\chi^2$ criteria for two degrees of freedom. The resulting confidence regions for the five NSI parameters are shown in Fig.~\ref{fig:Two_dim_NSI_plot1}. 

Let us briefly comment on the connection with Ref.~\cite{Li:2014mlo}, which investigated the sensitivity of JUNO to source and detector NSIs. That approach can be mapped onto the EFT framework using the expressions provided in Table X of Ref.~\cite{Falkowski:2019xoe}. Through this connection, one finds that the analysis of Ref.~\cite{Li:2014mlo} effectively corresponds to our $\im[\widetilde R]$ case. On the other hand, scalar and tensor interactions could not be covered in that study, as they introduce distinct energy dependences.

We close this section by briefly commenting on complementary constraints from other experiments. We recall that the combination of NSI coefficients constrained in our analysis is given by $[\widetilde X]=c_{23}[\epsilon_X]_{e\mu}-s_{23}[\epsilon_X]_{e\tau}$, for $X=R,S,T$. 
It is important to highlight that, because the NSIs considered here are flavor-violating, only (finite-distance) neutrino oscillation experiments exhibit a linear sensitivity to them. This is due to the fact that oscillations provide a non-zero baseline signal with which these flavor-violating NSIs can interfere. Conversely, non-oscillation experiments are only quadratically sensitive to such interactions. This quadratic dependence entails some degeneracies (e.g. the inability to disentangle the real and imaginary parts of the NSI parameters) and suppressed sensitivity. Nevertheless, this suppression is often heavily compensated for by the fact that non-oscillation searches do not rely on neutrino detection, allowing them to achieve significantly higher experimental precision. Finally, oscillation experiments operating at near-zero baselines functionally fall into the non-oscillation category regarding their NSI sensitivity.

Short-baseline reactor data constrain a different linear combination of $[\epsilon_X]_{e\mu}$ and $[\epsilon_X]_{e\tau}$, which also involves the PMNS CP phase. Assuming all relevant coefficients are of $\cO(1)$, the resulting bounds are typically at the level of $\cO(0.05-0.10)$~\cite{Falkowski:2019xoe}.

For a detailed discussion of the bounds from non-oscillation observables, we refer the reader to Refs.~\cite{Kopp:2025ffx,Falkowski:2019xoe,Falkowski:2021bkq}. Here, we summarize only the main conclusions. The specific bounds depend on the theoretical framework adopted. Within the LEFT framework, constraints on these NSIs can be extracted from high-precision low-energy observables, including nuclear $\beta$ decays, pion decays, and tests of CKM unitarity, as well as $\nu_\mu\to\nu_e$ searches at $L\approx 0$. These bounds are typically at the $\cO(10^{\eminus3}-10^{\eminus2})$ level. Conversely, if one assumes the SMEFT as the underlying high-energy framework, more stringent limits can be derived. High-$p_T$ searches at the LHC place bounds on scalar and tensor NSIs at the $\cO(10^{\eminus3})$ level. Even stronger constraints arise from charged lepton flavor-violating (cLFV) processes, which probe these interactions down to the $\cO(10^{\eminus4}-10^{\eminus8})$ level. As previously discussed, right-handed currents are not generated at dimension-6 within the SMEFT framework, see Eq.~\eqref{eq:eps_SMEFT_LEFT_matching}.

Lastly, prospects for constraining these coefficients with future data from FASER$\nu$~\cite{Falkowski:2021bkq}, DUNE~\cite{Kopp:2025ffx}, and muon-collider neutrino detectors~\cite{Kling:2025zsb} have also been studied, with projected sensitivities ranging from $\cO(1)$ - $\cO(10^{\eminus4})$, depending on the coupling and the assumptions.

%%%%%%%%%%%%%%%%%%%%%%%%%%%%%%%%%%%%%%%%%%%%%%%%%%%%%%%%%%%%%%%%%
%%%%%%%%%%%%%%%%%%%%%%%%%%%%%%%%%%%%%%%%%%%%%%%%%%%%%%%%%%%%%%%%%
%%%%%%%%%%%%%%%%%%%%%%%%%%%%%%%%%%%%%%%%%%%%%%%%%%%%%%%%%%%%%%%%%
%%%%%%%%%%%%%%%%%%%%%%%%%%%%%%%%%%%%%%%%%%%%%%%%%%%%%%%%%%%%%%%%%

%%%%%%%%%%%%%%%%%%%%%%%%%%%%%%%%%%%%%%%%%%%%%%%%%%%%%%%%%%%%%%%%%%%%%%%%%%%%%%%%%%%%
%%%%%%%%%%%%%%%%%%%%%%%%%%%%%%%%%%%%%%%%%%%%%%%%%%%%%%%%%%%%%%%%%%%%%%%%%%%%%%%%%%%%
\section{Conclusion}
\label{sec:conc}
%%%%%%%%%%%%%%%%%%%%%%%%%%%%%%%%%%%%%%%%%%%%%%%%%%%%%%%%%%%%%%%%%%%%%%%%%%%%%%%%%%%%
%%%%%%%%%%%%%%%%%%%%%%%%%%%%%%%%%%%%%%%%%%%%%%%%%%%%%%%%%%%%%%%%%%%%%%%%%%%%%%%%%%%%

We contribute to the systematic analysis of New Physics effects in neutrino experiments using Effective Field Theory (EFT) methods.
Our results can be summarized on two fronts.
On the theoretical side, we have shown that the lengthy QFT expression for the event rate in vacuum derived in Ref.~\cite{Falkowski:2019kfn}, Eq.~\eqref{eq:event-rate-formula}, can be written in a compact and  enlightening form in terms of matrix quantities with clear physical meaning, namely
\begin{align}
R_{\alpha\beta} = \tr\!\lzs F \frac{d\Phi^\alpha}{dE_\nu} F^\dagger \Sigma^\beta \dzs
=\frac{d\phi^\alpha}{dE_\nu}\,\tr\!\lzs \rho\, \Sigma^\beta \dzs
\,,
\end{align}
where $F$ is the evolution matrix, and $\Phi^\alpha$, $\Sigma^\beta$ and $\rho$ are generalized versions of the neutrino flux, cross-section and density matrix, respectively, defined in terms of QFT amplitudes. 
These quantities encode the creation, propagation and detection of the neutrino quantum state, capturing interference effects and oscillations in a basis independent form. 
We have discussed the extension of the QFT formalism to incorporate matter effects in the presence of generic interactions with left-handed neutrinos, finding that the standard QM treatment based on a Schr\"odinger-like equation is recovered, and that matter effects are trivially incorporated within our matrix framework through the evolution operator $F$. Finally, we have discussed the notion of oscillation probability in this general context, establishing that it is properly bounded between 0 and 1, and that it is in general \emph{not} a universal quantity---it depends on the details of the production and detection processes~\cite{Falkowski:2019kfn}.

On the phenomenological side, we have used this framework to derive analytical expressions for the oscillation observables relevant to medium-baseline reactor neutrino experiments such as JUNO, and have discussed how these generalize to longer-baseline configurations where matter effects become relevant. Together with Ref.~\cite{Falkowski:2019xoe}, which carried out an analogous analysis for short-baseline reactor neutrino experiments, this work completes the EFT treatment of this class of experiments. See also Ref.~\cite{Kopp:2025ffx} for a numerical study of EFT sensitivities at DUNE.

As a concrete application, we perform a first EFT analysis of the recent JUNO results~\cite{JUNO:2025gmd}. Interestingly, our study, which is admittedly simplified as it does not account for several refined experimental effects, is nonetheless capable of reproducing the JUNO results with reasonable accuracy, supporting its use for preliminary sensitivity studies such as the present one. We remark that we provide all details in App.~\ref{app:data} necessary to fully reproduce our analysis. We would like to encourage the JUNO collaboration to perform a full-fledged EFT analysis along these lines, using the analytical results obtained in this work and accounting for all relevant experimental effects, in the spirit of what was done by the Daya Bay collaboration~\cite{DayaBay:2024hya}. The expressions derived in this work also provide the necessary ingredients for a more ambitious joint EFT analysis of JUNO and Daya Bay data.

More broadly, this paper is part of a wider program aimed at performing a global EFT analysis of neutrino oscillation experiments (see, e.g., Ref.~\cite{Coloma:2023ixt}), where any given EFT interaction is treated consistently across experiments alongside the standard oscillation parameters---angles, phase, and mass splittings. This would allow one to track correlations among experiments and resolve flat or weakly bounded directions that appear in individual analyses, thereby shedding light on the UV implications of the current (dis)agreement among the various determinations of the PMNS parameters (e.g. $\theta_{12}$ from solar and reactor experiments). This global EFT program for neutrino data parallels analogous efforts in the quark flavor sector, where NP and CKM parameters must be analyzed simultaneously~\cite{Descotes-Genon:2018foz}, and would further allow one to study synergies between neutrino oscillation data and other sectors.
This was explored in Refs.~\cite{Breso-Pla:2023tnz,Terol-Calvo:2019vck,Coloma:2024ict}, where the interplay between neutrino data and electroweak precision observables was studied for flavor-conserving interactions. The present work represents a further contribution to this program.

%%%%%%%%%%%%%%%%%%%%%%%%%%%%%%%%%%%%%%%%%%%%%%%%%%%%%%%%%%%%%%%%%%%%%%%%%%%%%%%%%%%%
\section*{Acknowledgments}
%%%%%%%%%%%%%%%%%%%%%%%%%%%%%%%%%%%%%%%%%%%%%%%%%%%%%%%%%%%%%%%%%%%%%%%%%%%%%%%%%%%%
We thank Sergio Palomares-Ruiz for useful discussions. This work has been supported by MCIU/AEI/10.13039/501100011033 (grants CEX2023-001292-S and PID2023-146220NB-I00).

%%%%%%%%%%%%%%%%%%%%%%%%%%%%%%%%%%%%%%%%%%%%%%%%%%%%%%%%%%%%%%%%%%%%%%%%%%%%%%%%%%%%
%%%%%%%%%%%%%%%%%%%%%%%%%%%%%%%%%%%%%%%%%%%%%%%%%%%%%%%%%%%%%%%%%%%%%%%%%%%%%%%%%%%%
\appendix
%%%%%%%%%%%%%%%%%%%%%%%%%%%%%%%%%%%%%%%%%%%%%%%%%%%%%%%%%%%%%%%%%%%%%%%%%%%%%%%%%%%%
%%%%%%%%%%%%%%%%%%%%%%%%%%%%%%%%%%%%%%%%%%%%%%%%%%%%%%%%%%%%%%%%%%%%%%%%%%%%%%%%%%%%
\section{Consistency Properties of $\rho$ and $P_{\alpha\beta}$}
\addtocontents{toc}{\setcounter{tocdepth}{1}}
%%%%%%%%%%%%%%%%%%%%%%%%%%%%%%%%%%%%%%%%%%%%%%%%%%%%%%%%%%%%%%%%%%%%%%%%%%%%%%%%%%%%
%%%%%%%%%%%%%%%%%%%%%%%%%%%%%%%%%%%%%%%%%%%%%%%%%%%%%%%%%%%%%%%%%%%%%%%%%%%%%%%%%%%%
%%%%%%%%%%%%%%%%%%%%%%%%%%%%%%%%%%%%%%%%%%%%%%%%%%%%%%%%%%%%%%%%%%%%%%%%%%%%%%%%%%%%
%%%%%%%%%%%%%%%%%%%%%%%%%%%%%%%%%%%%%%%%%%%%%%%%%%%%%%%%%%%%%%%%%%%%%%%%%%%%%%%%%%%%
\subsection{Properties of $\rho$}
\label{app:rho}
%%%%%%%%%%%%%%%%%%%%%%%%%%%%%%%%%%%%%%%%%%%%%%%%%%%%%%%%%%%%%%%%%%%%%%%%%%%%%%%%%%%%
%%%%%%%%%%%%%%%%%%%%%%%%%%%%%%%%%%%%%%%%%%%%%%%%%%%%%%%%%%%%%%%%%%%%%%%%%%%%%%%%%%%%
We begin by recalling the QFT definition of the density matrix introduced in Eq.~\eqref{eq:rho_def}:
\begin{align}
    \rho=F \,\rho_0\,F^\dagger~,
    \qquad
    \rho_0=\frac{d\Phi^\alpha}{dE_\nu} \lzm \frac{d\phi^\alpha}{dE_\nu} \dzm^{\eminus1}~.
\end{align}
In the following, we demonstrate that $\rho$ satisfies the standard requirements of a quantum density matrix, namely Hermiticity, positivity, and normalization:
\begin{itemize}
    \item \textbf{Hermiticity:} Using that $d\phi^\alpha/dE_\nu\sim \int |\mathcal M^P|^2$ is real and positive, along with the fact that
    \begin{equation}
        \left[ \frac{d\Phi^\alpha}{dE_\nu} \right]_{\rho\eta}^*
        =\frac{N_S}{8\pi L^2 m_S}\int_{d\Pi_{P'}} \bar{\mathcal M}^P_{\alpha\rho}\mathcal M^P_{\alpha\eta}
        =\frac{N_S}{8\pi L^2 m_S}\int_{d\Pi_{P'}} \mathcal M^P_{\alpha\eta} \bar{\mathcal M}^P_{\alpha\rho}
        =\left[ \frac{d\Phi^\alpha}{dE_\nu} \right]_{\eta\rho}\,,
    \end{equation}
    it follows that $d\Phi^\alpha/dE_\nu$ is Hermitian. Therefore,
    \begin{equation}
        \rho^\dagger
        =\lzs F \frac{d\Phi^\alpha/dE_\nu}{d\phi^\alpha/dE_\nu}F^\dagger \dzs^\dagger
        =F\frac{(d\Phi^\alpha/dE_\nu)^\dagger}{d\phi^\alpha/dE_\nu}F^\dagger
        =F \frac{d\Phi^\alpha/dE_\nu}{d\phi^\alpha/dE_\nu}F^\dagger
        =\rho\,,
    \end{equation}
    which proves that $\rho$ is also Hermitian.
    \item \textbf{Positivity:} Positive semi-definiteness of $\rho$ follows directly from the linear-algebraic structure of the generalized flux. Up to overall constants and phase-space integration, $d\Phi^\alpha/dE_\nu$ is a rank-1 matrix, i.e., $d\Phi^\alpha/dE_\nu=P P^\dagger$, where $P_k\equiv {\cal M}^P_{\alpha k}$ is a $3\times 1$ vector in neutrino-family space. Hence $d\Phi^\alpha/dE_\nu$ is positive semidefinite since $v^\dagger P P^\dagger v=|P^\dagger v|^2\ge0$ for any vector $v$. Moreover, positive semidefiniteness is preserved under the map $A\mapsto F A F^\dagger$. Indeed, for any vector $v$
    \begin{equation}
    v^\dagger F A F^\dagger v=(F^\dagger v)^\dagger A (F^\dagger v)\ge 0\,,
    \end{equation}
    whenever $A$ is positive semidefinite. Applying this to $A=d\Phi^\alpha/dE_\nu$, and using $d\phi^\alpha/dE_\nu>0$, we obtain that $\rho$ is positive semidefinite.
    \item \textbf{Normalization ($\Tr\rho=1$):} Taking the trace of $\rho$, we obtain
    \begin{equation}
        \tr\rho 
        =\tr\lzs F \frac{d\Phi^\alpha/dE_\nu}{d\phi^\alpha/dE_\nu}F^\dagger \dzs
        =\tr\lzs F^\dagger F \frac{d\Phi^\alpha/dE_\nu}{d\phi^\alpha/dE_\nu} \dzs
        =\tr\lzs \frac{d\Phi^\alpha/dE_\nu}{d\phi^\alpha/dE_\nu} \dzs=1\,,
    \end{equation}
    where we use the cyclicity of the trace and the unitarity of $F$. Therefore, this proves that $\rho$ is properly normalized.
\end{itemize}
We note that, since $d\Phi^\alpha/dE_\nu$ is a positive-semidefinite rank-one matrix, there always exists a basis in which only a single diagonal entry is nonzero. This basis may be interpreted as the production flavor basis associated with the given process. An analogous statement holds for the generalized cross section, defining a corresponding detection flavor basis. In general, these two bases are process-dependent and need not coincide. In the SM, however, both reduce to the standard flavor basis defined by the weak interactions. 

Finally, the properties listed above imply that the eigenvalues $\lambda_i$ of a density matrix are real, non-negative, and satisfy $\sum_i\lambda_i=1$. Therefore, $\Tr\rho^2=\sum_i \lambda_i^2\le \big(\sum_i \lambda_i\big)^2=1$, which establishes the purity bound $\Tr\rho^2\le 1$. Note that, within our framework, one has $\Tr\rho^2=1$, which follows from the fact that $d\Phi^\alpha/dE_\nu$ has a single non-zero diagonal entry, as mentioned above.  Physically, this reflects that the neutrino state is pure by construction, since decoherence effects are not included in our framework.

%%%%%%%%%%%%%%%%%%%%%%%%%%%%%%%%%%%%%%%%%%%%%%%%%%%%%%%%%%%%%%%%%%%%%%%%%%%%%%%%%%%%
%%%%%%%%%%%%%%%%%%%%%%%%%%%%%%%%%%%%%%%%%%%%%%%%%%%%%%%%%%%%%%%%%%%%%%%%%%%%%%%%%%%%
\subsection{Properties of $P_{\alpha\beta}$}
\label{sec:proof}
%%%%%%%%%%%%%%%%%%%%%%%%%%%%%%%%%%%%%%%%%%%%%%%%%%%%%%%%%%%%%%%%%%%%%%%%%%%%%%%%%%%%
%%%%%%%%%%%%%%%%%%%%%%%%%%%%%%%%%%%%%%%%%%%%%%%%%%%%%%%%%%%%%%%%%%%%%%%%%%%%%%%%%%%%
Let us show that the probability defined as
\begin{equation}
P_{\alpha\beta}
~=~ \frac{\Tr\left[ \rho\,\Sigma^\beta \right]}{\Tr\left[ \Sigma^\beta \right]}~,
\end{equation}
is bounded between zero and one. Much like the flux, the generalized cross section is, up to overall constants and phase-space integration, a positive-semidefinite rank-one matrix, i.e., $\Sigma^\beta=D D^\dagger$,
where $D_k\equiv \bar{\cal M}^D_{\beta k}$ is a $3\times 1$ vector in neutrino-family space. The probability can therefore be written, using the cyclicity of the trace, as 
\begin{equation}
    P_{\alpha\beta}=\frac{\tr[\rho \,DD^\dagger]}{\tr[DD^\dagger]}=\frac{\tr[D^\dagger\rho\,D]}{\tr[D^\dagger D]}=\frac{D^\dagger \rho\,D}{D^\dagger D}\,.
\end{equation}
Since $\rho$ is positive semidefinite (see App.~\ref{app:rho}), one has $D^\dagger \rho \,D \ge 0$, which immediately implies $P_{\alpha\beta}\ge 0$. Moreover, because $\rho$ is Hermitian, positive semidefinite, and normalized, all of its eigenvalues satisfy $0\le \lambda_i\le 1$. Therefore, the matrix $\mathbb{1}-\rho$ is positive semidefinite, implying $D^\dagger (\mathbb{1}-\rho) D \ge 0$ or, equivalently, $D^\dagger \rho\,D \le D^\dagger D$. It then follows that 
\begin{equation}
    P_{\alpha\beta}= \frac{D^\dagger \rho\,D}{D^\dagger D}\le 1\,,
\end{equation}
which proves that $0\le P_{\alpha\beta}\le 1$.

%%%%%%%%%%%%%%%%%%%%%%%%%%%%%%%%%%%%%%%%%%%%%%%%%%%%%%%%
%%%%%%%%%%%%%%%%%%%%%%%%%%%%%%%%%%%%%%%%%%%%%%%%%%%%%%%%
\clearpage
\section{Digitized Data Used for the Analysis}
\label{app:data}
%%%%%%%%%%%%%%%%%%%%%%%%%%%%%%%%%%%%%%%%%%%%%%%%%%%%%%%%
%%%%%%%%%%%%%%%%%%%%%%%%%%%%%%%%%%%%%%%%%%%%%%%%%%%%%%%%
In the absence of publicly available JUNO data, the spectra used in this analysis are digitized and provided for transparency and reproducibility, and should not be interpreted as official JUNO data. We retain 65 of the 66 original points, as the final bin lies at the extreme tail and cannot be reliably digitized.

%%%%%%%%%%%%%%%%%%%%%%%%%%%%%%%%%%%%%%%%%%%%%%%%%%%%%%%%
%%%%%%%%%%%%%%%%%%%%%%%%%%%%%%%%%%%%%%%%%%%%%%%%%%%%%%%%
\begin{table}[!htbp]
    \centering

\renewcommand{\arraystretch}{1.05}

{\scalebox{0.95}{
\begin{tabular}{|c|c|c|c|c|c|c|c|c|}

\hline

$i$& $E_{\sscript{pr}}^i$& $N^{\sscript{S+B}}_i$& $N^{\sscript{unosc}}_i$& $N^{\sscript{Geo}}_i$& $N^{\sscript{LiHe}}_i$& $N^{\sscript{BiPo}}_i$& $N^{\sscript{World}}_i$& $N^{\sscript{Other}}_i$\\

\hline

1& 0.84&  7&   7.5&  0.7& 1.5& 0.1& 0.5& 0.2\\

2& 1.04& 25&  34.9&  7.6& 1.5& 0.2& 0.5& 0.2\\

3& 1.13& 48&  50.1& 10.2& 1.5& 0.2& 0.7& 0.2\\

4& 1.23& 54&  62.7& 12.1& 1.5& 0.2& 0.7& 0.3\\

5& 1.32& 45&  66.0& 10.0& 1.5& 0.2& 0.8& 0.3\\

6& 1.44& 37&  74.2&  3.5& 1.8& 0.2& 0.7& 0.3\\

7& 1.54& 29&  87.4&  2.7& 1.8& 0.3& 0.8& 0.2\\

8& 1.64& 33&  98.3&  2.7& 2.1& 0.2& 0.9& 0.3\\

9& 1.74& 32& 109.6&  2.7& 2.4& 0.2& 0.9& 0.3\\

10& 1.84& 30& 120.5&  3.1& 2.4& 0.3& 1.2& 0.2\\

11& 1.94& 40& 129.1&  3.3& 2.7& 0.3& 1.2& 0.2\\

12& 2.03& 43& 137.7&  3.3& 2.4& 0.3& 1.2& 0.2\\

13& 2.14& 36& 145.3&  3.5& 3.0& 0.3& 1.1& 0.4\\

14& 2.23& 49& 152.9&  3.1& 3.0& 0.3& 1.2& 0.8\\

15& 2.33& 29& 159.5&  2.7& 2.7& 0.2& 0.9& 0.2\\

16& 2.44& 43& 164.8&  0.0& 3.0& 0.3& 1.1& 0.1\\

17& 2.53& 30& 171.0&  0.0& 3.3& 0.3& 1.1& 0.1\\

18& 2.63& 38& 175.3&  0.0& 3.3& 0.3& 0.9& 0.2\\

19& 2.73& 49& 176.3&  0.0& 3.8& 0.3& 1.1& 0.2\\

20& 2.83& 38& 176.6&  0.0& 3.6& 0.3& 1.1& 0.2\\

21& 2.94& 39& 177.0&  0.0& 3.8& 0.2& 0.8& 0.2\\

22& 3.04& 36& 175.7&  0.0& 4.1& 0.2& 1.1& 0.2\\

23& 3.14& 39& 175.7&  0.0& 4.4& 0.3& 1.1& 0.2\\

24& 3.24& 45& 174.3&  0.0& 4.4& 0.2& 1.1& 0.2\\

25& 3.33& 41& 172.4&  0.0& 4.4& 0.3& 1.1& 0.2\\

26& 3.43& 52& 169.0&  0.0& 4.4& 0.4& 0.9& 0.2\\

27& 3.54& 52& 164.8&  0.0& 4.7& 0.4& 0.8& 0.2\\

28& 3.64& 42& 159.8&  0.0& 4.7& 0.3& 0.8& 0.2\\

29& 3.75& 62& 153.2&  0.0& 4.4& 0.3& 0.8& 0.2\\

30& 3.84& 43& 145.3&  0.0& 4.7& 0.4& 0.7& 0.2\\

\hline

\end{tabular}
}}
    \caption{Bin by bin numerical values extracted from Fig.~3 of Ref.~\cite{JUNO:2025gmd} for the prompt energy $E_{\sscript{pr}}^i$, the total number of events $N^{\sscript{S+B}}_i$, the unoscillated reactor neutrino expectation $N^{\sscript{unosc}}_i$, and the backgrounds $N^{\sscript{B}}_i$ (B = Geo, LiHe, BiPo, World, Other). The $N_i^B$ values in the table are obtained after normalizing them to their corresponding pre-fit values shown in Table 1 of Ref.~\cite{JUNO:2025gmd}.}
    \label{tab:data-1}
\end{table}
%%%%%%%%%%%%%%%%%%%%%%%%%%%%%%%%%%%%%%%%%%%%%%%%%%%%%%%%
%%%%%%%%%%%%%%%%%%%%%%%%%%%%%%%%%%%%%%%%%%%%%%%%%%%%%%%%
\FloatBarrier

%%%%%%%%%%%%%%%%%%%%%%%%%%%%%%%%%%%%%%%%%%%%%%%%%%%%%%%%
%%%%%%%%%%%%%%%%%%%%%%%%%%%%%%%%%%%%%%%%%%%%%%%%%%%%%%%%
\begin{table}[!htbp]
    \centering

\renewcommand{\arraystretch}{1.05}

{\scalebox{0.95}{
\begin{tabular}{|c|c|c|c|c|c|c|c|c|}

\hline

$i$& $E_{\sscript{pr}}^i$& $N^{\sscript{S+B}}_i$& $N^{\sscript{unosc}}_i$& $N^{\sscript{Geo}}_i$& $N^{\sscript{LiHe}}_i$& $N^{\sscript{BiPo}}_i$& $N^{\sscript{World}}_i$& $N^{\sscript{Other}}_i$\\

\hline

31& 3.94& 38& 137.3&  0.0& 5.0& 0.4& 0.8& 0.2\\

32& 4.04& 47& 132.4&  0.0& 4.7& 0.3& 0.9& 0.2\\

33& 4.14& 44& 128.4& 0.0& 5.0& 0.2& 0.9& 0.2\\

34& 4.24& 53& 124.8& 0.0& 4.7& 0.3& 0.7& 0.1\\

35& 4.35& 58& 121.8& 0.0& 4.7& 0.0& 0.7& 0.2\\

36& 4.44& 48& 114.2& 0.0& 5.0& 0.1& 0.7& 0.2\\

37& 4.53& 44& 113.5& 0.0& 5.3& 0.1& 0.7& 0.2\\

38& 4.64& 65& 109.2& 0.0& 5.3& 0.1& 0.8& 0.1\\

39& 4.74& 49& 104.9& 0.0& 5.0& 0.2& 0.8& 0.2\\

40& 4.84& 48&  99.7& 0.0& 5.0& 0.2& 0.8& 0.3\\

41& 4.94& 40&  94.7& 0.0& 4.7& 0.0& 0.8& 0.3\\

42& 5.04& 43&  88.1& 0.0& 5.0& 0.0& 0.8& 0.3\\

43& 5.15& 49&  83.1& 0.0& 5.0& 0.1& 0.8& 0.2\\

44& 5.25& 54&  79.2& 0.0& 5.3& 0.1& 0.8& 0.1\\

45& 5.34& 34&  74.6& 0.0& 5.0& 0.1& 0.8& 0.1\\

46& 5.44& 30&  64.0& 0.0& 4.7& 0.1& 0.9& 0.2\\

47& 5.54& 33&  59.0& 0.0& 5.0& 0.2& 0.7& 0.2\\

48& 5.64& 30&  54.4& 0.0& 4.7& 0.2& 0.8& 0.2\\

49& 5.75& 33&  51.8& 0.0& 4.7& 0.2& 0.8& 0.2\\

50& 5.85& 37&  47.8& 0.0& 5.3& 0.0& 0.5& 0.2\\

51& 5.95& 33&  44.5& 0.0& 4.7& 0.0& 0.5& 0.1\\

52& 6.05& 26&  41.5& 0.0& 4.7& 0.0& 0.5& 0.1\\

53& 6.15& 23&  39.5& 0.0& 4.7& 0.0& 0.5& 0.1\\

54& 6.24& 18&  36.6& 0.0& 4.7& 0.0& 0.5& 0.1\\

55& 6.34& 19&  31.3& 0.0& 4.4& 0.0& 0.5& 0.2\\

56& 6.45& 21&  28.0& 0.0& 4.4& 0.0& 0.7& 0.2\\

57& 6.54& 11&  24.3& 0.0& 4.4& 0.0& 0.7& 0.2\\

58& 6.69& 12&  18.4& 0.0& 4.4& 0.0& 0.5& 0.2\\

59& 6.90& 10&  17.4& 0.0& 4.4& 0.0& 0.5& 0.3\\

60& 7.10&  9&  13.1& 0.0& 4.1& 0.0& 0.5& 0.3\\

61& 7.28&  7&  10.1& 0.0& 4.1& 0.0& 0.5& 0.2\\

62& 7.55&  4&   7.8& 0.0& 3.8& 0.0& 0.5& 0.1\\

63& 7.90&  4&   4.5& 0.0& 3.3& 0.0& 0.5& 0.0\\

64& 8.35&  3&   3.2& 0.0& 2.7& 0.0& 0.5& 0.1\\

65& 9.00&  2&   1.9& 0.0& 2.7& 0.0& 0.8& 0.1\\

\hline

\end{tabular}
}}
    \caption{Table~\ref{tab:data-1} continued.}
    \label{tab:data-2}
\end{table}
%%%%%%%%%%%%%%%%%%%%%%%%%%%%%%%%%%%%%%%%%%%%%%%%%%%%%%%%
%%%%%%%%%%%%%%%%%%%%%%%%%%%%%%%%%%%%%%%%%%%%%%%%%%%%%%%%

%\clearpage
\bibliographystyle{JHEP}
\bibliography{References}

\end{document}